\documentclass[twocolumn]{aastex6}

\newcommand{\Hiiregs}{\ion{H}{2} regions}
\newcommand{\tsqd}{$t^2$}
\newcommand{\temp}{$T_{\rm e}$}
\newcommand{\dens}{$n_{\rm e}$}

\shorttitle{Nebular spectroscopy}
\shortauthors{Peimbert et al.}

\begin{document}

\title{Nebular Spectroscopy: a guide on \ion{H}{2} regions and Planetary Nebulae}
\author{Manuel Peimbert, Antonio Peimbert, and Gloria Delgado-Inglada}
\affil{Instituto de Astronom\'{i}a \\
Universidad Nacional Aut\'onoma de M\'exico\\
Apartado Postal 70-264, C.P. 04510\\ 
M\'exico, CDMX, Mexico}

\begin{abstract}
We present a tutorial on the determination of the physical conditions and chemical abundances in gaseous nebulae. We also include a brief review of recent results on the study of gaseous nebulae, their relevance for the study of stellar evolution, galactic chemical evolution, and  the evolution of the universe. One of the most important problems in abundance determinations is the existence of a discrepancy between the abundances determined with collisionally excited lines and those determined by recombination lines, this is called the ADF (abundance discrepancy factor) problem; we review results related to this problem. Finally, we discuss possible reasons for the large $t^2$ values observed in gaseous nebulae.
\end{abstract}


\section{Introduction}\label{sec:intro}

There are two main types of gaseous nebulae that we will discuss in this review: \Hiiregs\ and planetary nebulae (PNe). Precise models of individual nebulae permit us to determine accurate abundances and the abundances allow us to test models of stellar evolution, galactic chemical evolution, and the evolution of the Universe as a whole. 

\Hiiregs\ are sites where star formation is occurring; therefore they provide us with the initial abundances from which stars are made at present, abundances are paramount to test models of the chemical evolution of galaxies. For spiral galaxies they provide us with radial abundance gradients of heavy elements relative to hydrogen, that have to be explained by models of galactic chemical evolution \citep{Berg2013, Sanchez2014, Bresolin2015, Esteban2015, Magrini2016, ArellanoCordova2016}. Irregular galaxies, that have a high fraction of their mass in the form of gas, where almost no heavy elements are present, permit us to determine the primordial abundance of helium and hydrogen due to Big-Bang nucleosynthesis \citep{Ferland2010, Izotov2014, Aver2015, Peimbert2016}.

PNe are produced by intermediate mass stars, those in the 0.8 M$_{\odot}$ to 8 M$_{\odot}$ range, while they transit from the red giant stage to the white dwarf stage. The study of these nebulae show that intermediate mass stars are responsible for most of the nitrogen, about half the carbon, and a non negligible fraction of oxygen and helium present in the universe \citep[e.g.][]{Karakas2014}. These stars also produce an important fraction of the slow neutron process elements, like rubidium, strontium, yttrium, zirconium, cesium, barium, lanthanum, and praseodymium. Big-Bang nucleosynthesis created all the hydrogen and deuterium, most of the helium  and a fraction of the lithium. Massive stars, those with more than 8 M$_{\odot}$, produce most of the remaining elements including the fast  neutron process elements as well as a fraction of the helium, carbon, nitrogen, and oxygen abundances \citep[e.g.][]{Pagel2009}.

The classical textbook on the study of gaseous nebulae and active galactic nuclei is that by \citet{Osterbrock2006}. This book discusses in depth the physical processes in PNe and \Hiiregs. It also mentions previous books and many review papers that present earlier ideas and results on these objects. Other texts that, in the past,  were considered fundamental in the study of photoionized regions are those by \citet{Stromgren1939, Stromgren1948}, \citet{Seaton1960}, \citet{Osterbrock1974,Osterbrock1989}, and \citet{Aller1984}. 

Other relevant review articles and books related to the physical conditions of gaseous nebulae are: a) \citet{Kwok2000} on the origin and evolution of planetary nebulae, b) \citet{Ferland2003} on quantitative spectroscopy of photoionized clouds, c) \citet{Dopita2003} on the astrophysics of the ionized universe, d) \citet{Stasinska2004} on cosmochemistry the melting pot of the elements, e) \citet{Pagel2009} on nucleosynthesis and the chemical evolution of galaxies, f) \citet{Stasinska2012} on oxygen in the universe.

In the last few years there have been four meetings dedicated only to planetary nebulae, two of the symposia organized by the IAU \citep{Manchado2012, Liu2017}, and two conferences on asymmetric planetary nebulae \citep{Zijlstra2010, Morisset2014}. The IAU Planetary Nebula Working Group produced a review paper on the present and future of PNe and their central stars research and related subjects \citep{Kwitter2014}.

\citet{PerezMontero2017} recently published a tutorial which deals specifically with the determination of chemical abundances of  \Hiiregs\ through the so-called direct method.

In this paper we present a combination of a tutorial and a review paper on recent results related to the study of \Hiiregs\ and PNe. Section 2 describes some basic physical processes present in gaseous nebulae. Section 3 describes the main methods used to determine the physical conditions in gaseous nebulae. Section 4 discusses  methods to determine the total abundances of the elements in gaseous nebulae, with special emphasis on the determination of the ionization correction factors to take into account the ions that are not observed. Section 5 presents recent results derived from the study of galactic and extragalactic  \Hiiregs. Section 6 presents recent results derived from planetary nebulae. Section 7 includes a general discussion on possible physical reasons of why the temperature inhomogeneities are considerably higher than those predicted by photoionization models of chemically homogeneous nebulae. Some final remarks are discussed in Section 8.  

\section{Brief discussion of physical processes in gaseous nebulae} \label{sec:processes}

There are several processes occurring in the ionized gas of planetary nebulae and \Hiiregs. Our aim in this section is to briefly explain some concepts (such as photoionization, recombination, heating, cooling, and emission line mechanisms) that are necessary to understand the subsequent sections. It is beyond the scope of this paper to describe in detail all of them and we suggest the reader to review the books and papers mentioned above (in particular, the book by \citealt{Osterbrock2006}) to understand the physics involved in these objects. 

\subsection{Photoionization and recombination processes: local ionization equilibrium}\label{ssec:ionization_equilibrium}

The basis of the study of photoionized regions is to assume an equilibrium between ionization and recombination. It has long been known that, in equilibrium, a photoionized region will have a large volume of ionized gas surrounded by a relatively narrow transition zone, where the gas goes from being nearly completely ionized to nearly completely neutral  \citep[e.g.][]{Stromgren1939,Osterbrock2006}. 

Since approximately 90\% of the atoms of the ISM are hydrogen, to a first approximation the equilibrium is studied for a gas made up entirely of hydrogen atoms. Locally one must study the equilibrium between recombination and ionization
\begin{equation}
n_e  n_p \alpha_A({\rm H}^0, T_e)=
n({{\rm H}^0})\int_{\nu_0}^{\infty}{{4 \pi J_{\nu}}\over{h\nu}} a_\nu d\nu,
\end{equation}
where $n_e$, $n_p$, $n({{\rm H}^0})$, represent the electron, proton, and neutral hydrogen density, $\alpha_A({\rm H}^0, T_e)$ represents the recombination coefficient for hydrogen at a given temperature, $h\nu_0$ the energy required to ionize hydrogen (13.6 eV), $J_{\nu}$ represents the local radiation, and $a_\nu$ the ionization cross section of a given photon. In photoionized regions $\int_{\nu_0}^{\infty}{{4 \pi J_{\nu}}\over{h\nu}} a_\nu d\nu$ dominates over $ n_e\alpha({\rm H}^0, T_e)$ and the ionization fraction is nearly one for most of the volume.

Globally one can estimate the maximum volume that can be ionized by a constant ionization source. The first thing to notice is that recombinations to the ground level will produce photons with energy greater that $h\nu_0$, thus they can be considered an additional source of ionization; this source usually accounts for approximately 40\% of the ionizating photons. One solution to avoid this problem is to ignore recombinations to the ground level (since the subsequent ionization will cancel out the recombination). The recombination rate to all levels but the ground level is usually referred as $\alpha_B({\rm H}^0, T_e)$. Since there will be a steady rate of recombinations, ionizing photons will be required to keep a volume of gas ionized; those photons will be exhausted when a volume of size ${{4 \pi}\over3} r_S^3$ is ionized, where $r_S$ (the Str\"omgren radius) can be estimated as
\begin{equation}
Q({\rm H}^0)={{4 \pi}\over3}r_S^3n^2({\rm H})\alpha_B({\rm H}^0, T_e),
\end{equation}
here $Q({\rm H}^0)$ represents the rate of ionizing photons produced by the central star, and $n({\rm H}$) the total hydrogen density.

For nebulae with realistic chemical compositions, the ionization of helium needs to be modeled as well as those of heavier elements. In general neither will affect much the results for hydrogen, since helium  recombination will, in general, return hydrogen   ionizing photons, and the abundance of heavy elements is very small. However, the degree of ionization of heavy elements can not be neglected, since it turns out to be very important for the temperature equilibrium.

\subsection{Heating and cooling: local thermal equilibrium}\label{ssec:thermal_equilibrium}

The previous results cannot be properly determined without knowing the physical conditions of the gas of the nebula. While the density is generally considered to be given by the characteristics of the gas previous to the photoionization, the energetics of photoionized regions is dominated by photoionization and the temperature strongly depends on the physics of photoionization. As such, in equilibrium, the temperature of the gas comes from a balance of local cooling and heating processes.

Heating will come from photoionizing photons, and hydrogen photoionization will account for at least 90\% of this heating. Helium photoionization will account for 10\%, or less, of the total heating, while photoionization of heavy elements will only produce trace amounts of heating. Additional heating can come from dust photo-heating, or from free-free absorption, but neither of these will dominate the energetics. 

The heating due to photoionization will be due to the excess energy of photo ionizing photons beyond the ionizing threshold of the different atoms and ions (13.6 eV for hydrogen) consequently it will be proportional to the number of photoionizations and to the temperature of the photoionizing star. On the other hand, in equilibrium, the number of photoionizations is equal to the number of recombinations, which in turn is proportional to the density squared for most of the object.

Other possible sources of heating are cosmic rays and shock waves, these are traditionally considered to be unimportant, and are thus ignored. But, while the former is not expected to be important, the latter can, locally, contribute with an important fraction of the heating, see Section \ref{sec:t2_large}.

To balance the heating, one must consider the possible sources of cooling. The most obvious source that must be considered is recombination where the kinetic energy of the captured photon is removed from the gas. This is however not the most important source of cooling. 

If one were to consider a gas made up entirely of hydrogen   (or hydrogen and helium) recombination would indeed be the best way to remove energy from the nebula. In this scenario the heating is proportional to the temperature of the star, and the cooling is proportional to the temperature of the nebula, so balance will occur when the gas has a temperature similar to that of the star (in fact, when done carefully, the gas ends up being hotter than the ionizing star). This is not seen in nature: ionizing stars of \Hiiregs\ have temperatures of $T_{star}\sim30000-45000$ K, while \Hiiregs\ often have temperatures of $T_e\sim7000-15000$ K; central stars of PNe have temperatures of $T_{star}\sim30000-200000$ K, while PNe often have temperatures of $T_e\sim7000-20000$ K. The difference is due to the additional cooling produced by forbidden lines excited by collisions.\\

\subsection{Emission line mechanisms}\label{ssec:el_mechanisms}

\begin{figure}[ht!]
\centering
\includegraphics[trim = 20 0 15 0, clip, width = 8.5cm]{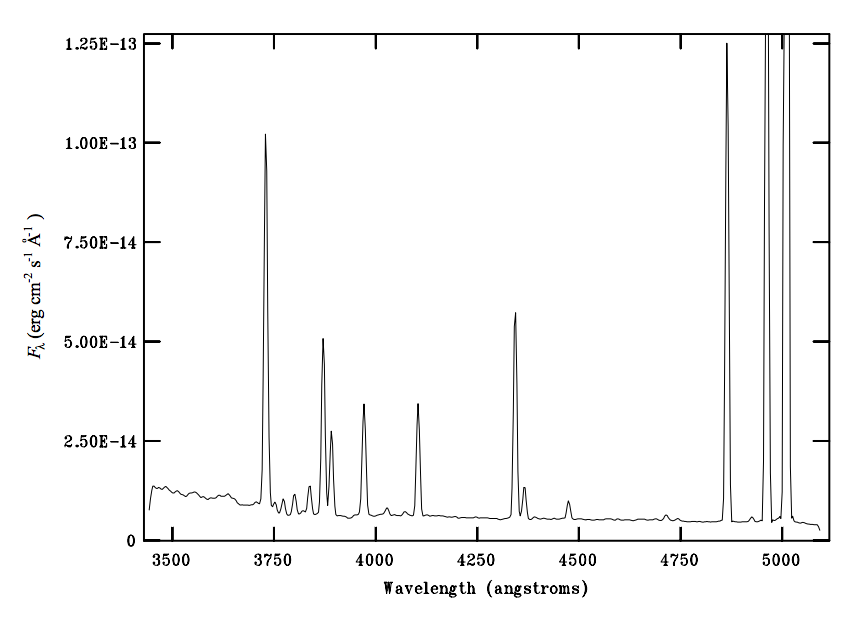}
\includegraphics[trim = 20 0 15 0, clip, width = 8.5cm]{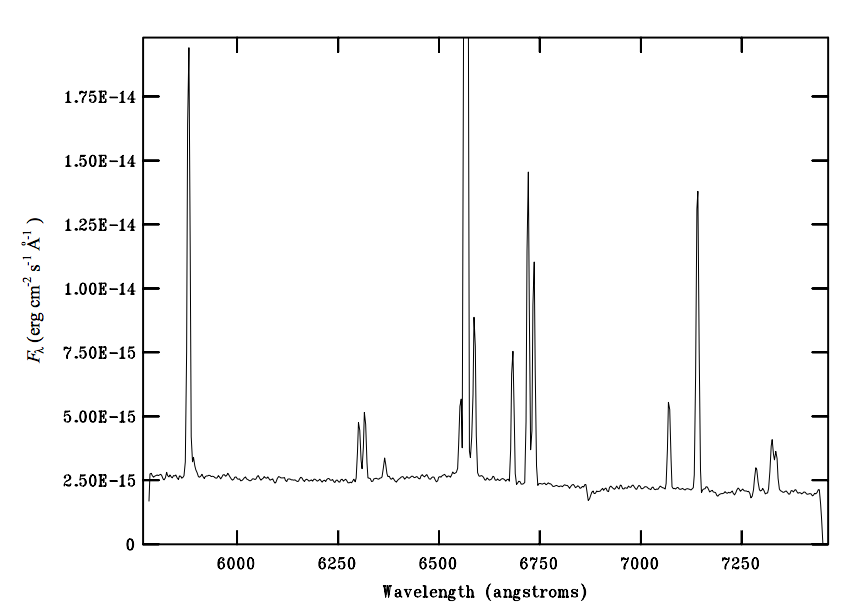}
\caption{Blue and red spectra for region A of NGC~346 obtained at CTIO with the 4 m telescope and the R-C spectrograph. The figures have been taken from \citet{Peimbert2000}.\label{fig:spectrangc346}}
\end{figure}

The spectra of ionized nebulae are characterized by a weak continuum and strong emission lines. Figure~\ref{fig:spectrangc346} shows some of the most conspicuos emission lines: [\ion{O}{2}] $\lambda$3727, $\lambda$3729, H$\delta$, H$\gamma$, H$\beta$,  [\ion{O}{3}] $\lambda$4363, $\lambda$4959, $\lambda$5007 in the blue part of the spectrum; and H$\alpha$, [\ion{N}{2}] $\lambda$6548, $\lambda$6583, He I $\lambda$6678, [\ion{S}{2}] $\lambda$6717, $\lambda$6731, He I $\lambda$7065, [\ion{Ar}{3}] $\lambda$7135 in the red part of the spectrum.

These lines are produced when an atom (or ion) is de-excited after being excited by photons or by collisions between atoms or atoms and electrons. The weak continuum is due to bound-free, free-free, two photon emission, dust scattered light, and starlight for extragalactic \Hiiregs. The main mechanisms producing emission lines are recombination, collisional excitation, and photoexcitation. One particular line can be produced by several mechanisms but usually one of them dominates over the others. Most of the strongest lines in ionized nebulae are produced by collisional excitation. 

\begin{figure}[ht!]
\centering
\includegraphics[trim = 20 0 15 0, clip, width = 8.5cm]{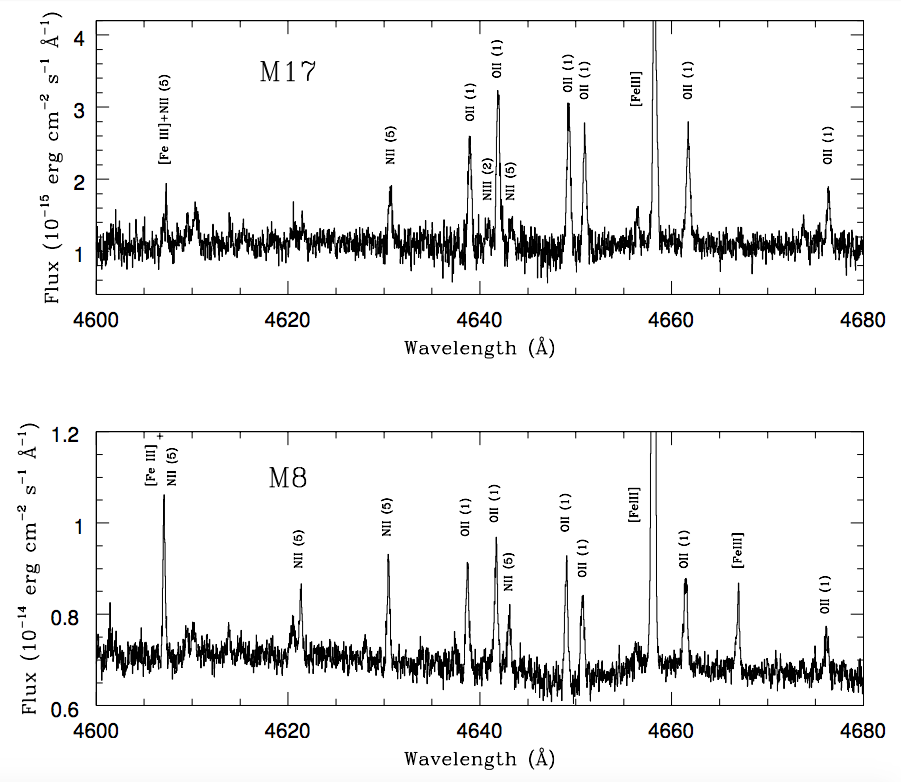}
\caption{Sections of the echelle spectra of the Galactic \Hiiregs\ M8 and M17 obtained with the high resolution spectrograph UVES at the Very Large Telescope. The figure has been taken from \citet{GarciaRojas2007}.\label{fig:spectraHII}}
\end{figure}

\begin{figure}[ht!]
\centering
\includegraphics[trim = 20 0 20 0, clip, width = 8.5cm]{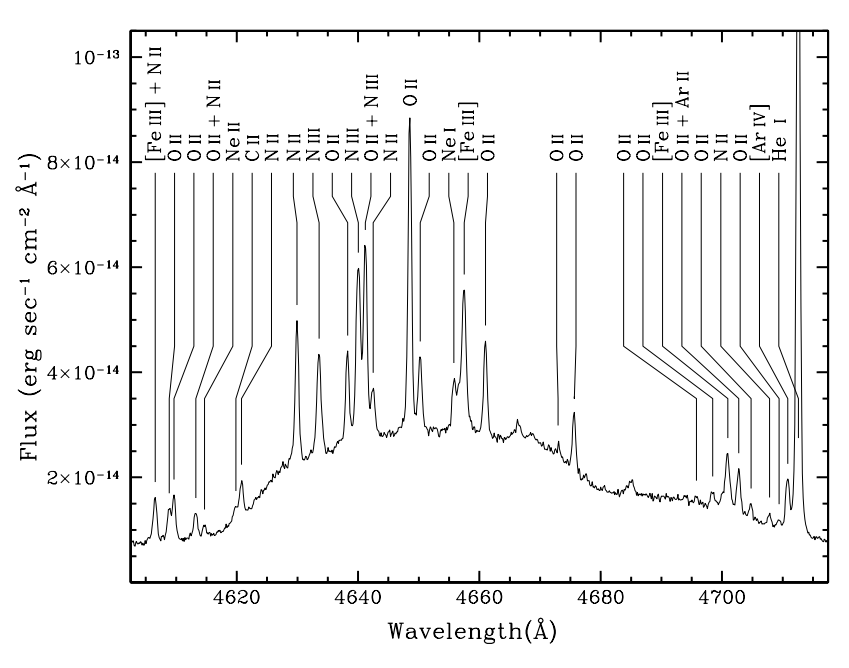}
\caption{Sections of the echelle spectrum of the Galactic PN NGC~5315 obtained with the high resolution spectrograph UVES at the Very Large Telescope. The broad feature is due to the [WC4] Wolf-Rayet type central star. The figure has been taken from \citet{Peimbert2004}.\label{fig:spectraPN}}
\end{figure}

Deeper spectra allow to measure the considerable fainter recombination lines. Figures \ref{fig:spectraHII} and \ref{fig:spectraPN} show three published echelle spectra of the \Hiiregs\ M8 and M17 obtained by \citet{GarciaRojas2007} and the PN NGC~5315 obtained by \citet{Peimbert2004}. These portions of the spectra show several recombination lines (such as \ion{C}{2}, \ion{O}{2}, \ion{N}{2}, \ion{Ne}{2}, and \ion{N}{3} lines) together with some collisonally excited lines of [\ion{Fe}{3}] and [\ion{Ar}{4}]. The broad emission feature in the spectra of NGC~5315 is due to the Wolf-Rayet type central star.

In general, the intensity of an emission line, $I(\lambda)$, can be written as:
\begin{equation}
\label{equ:emiss}
I(\lambda) = \int j_\lambda ds = \int n(X^{+i}) n_e \epsilon_\lambda(T_e)ds,
\end{equation}
where $j_\lambda$ is the emission coefficient, $n(X^{+i})$ is the density of the ion that emits the line, $n_{\rm e}$ is the electron density, and $\epsilon_\lambda$ is the emissivity.  For example, the H$\beta$ and [\ion{O}{3}] $\lambda5007$ intensity lines are given by:
\begin{equation}
I({\rm H}\beta) =  \int n({\rm H}^{+}) n_e \epsilon({{\rm H}\beta}, T_e)ds,
\end{equation}
\begin{equation}
I([{\rm O\,III}]) = \int n({\rm O}^{++}) n_e \epsilon(\lambda5007, T_e)ds.
\end{equation}
The emissivities of recombination lines (RLs) and collisionally excited lines (CELs) are discussed below.

 \subsubsection{Recombination lines}\label{sssec:elm_recombination}

RLs are produced when free electrons are captured by ions and descend from excited to lower levels emitting photons in the process. These lines are also known as permitted lines since they typically satisfy all the selection rules for an electric dipole transition. Most of the bright RLs in emission spectra are from hydrogen and helium. Metals, such as carbon, nitrogen, and oxygen, also produce RLs but are much weaker due to their lower abundances with respect to hydrogen and helium. 

The emission coefficient of a recombination line, $j_{\rm {nn'}}$, is given by: 
\begin{equation}
j_{nn'} = \frac{h\nu_{nn'}}{4\pi} n(X^{+i}) n_{e}\alpha^{\it eff}_{nn'},
\end{equation}
where $h\nu_{nn'}$ is the energy difference between the two levels and $\alpha^{\it eff}_{nn'}$ represents the effective recombination coefficient. 

Some of the RLs that can be found in the spectra of ionized nebulae are: \ion{H}{1} lines (e.g., H$\alpha$ at 6563 \AA, H$\beta$ at 4861 \AA, H$\gamma$ at 4340 \AA), \ion{He}{1} lines (e.g., 5875 and 4471 \AA), \ion{He}{2} lines (e.g., 4686 \AA), \ion{O}{1} lines (e.g., 8446 and 8447 \AA), \ion{O}{2} lines (e.g., 4639, 4642, 4649 \AA), \ion{O}{3} lines (e.g., 3265 \AA), \ion{O}{4} lines (e.g., 4631 \AA), \ion{C}{2} lines (e.g., 4267 \AA), \ion{C}{3} lines (e.g., 4647 \AA), \ion{C}{4} lines (e.g., 4657 \AA), \ion{N}{2} lines (e.g., 4237 and 4242 \AA), \ion{N}{3} lines (e.g., 4379 \AA), \ion{N}{4} lines (e.g., 4606 \AA), and \ion{Ne}{2} lines (e.g., 3694 \AA).

The \ion{H}{1} photons emitted by recombination may escape or not. In an optically thin nebula, all the emitted photons will escape; this is known as Case A \citep{Baker1938}. On the other hand, in an optically thick nebula, all the hydrogen   Lyman photons will be absorbed; this is called Case B. The intermediate situations between these two extreme cases are explained below.\\

\subsubsection{Collisionally excited lines}\label{sssec:elm_collisions}

In contrast with hydrogen and helium, the energies of the first excited levels of some ions of heavy elements are within a few eV of the ground level, which makes relatively easy to reach them through collisions with electrons. CELs are produced when the atoms, excited through collisions, decay via radiative transitions. Although the transition probability of these lines is low, the relatively low density of ionized nebulae, allows these transitions to occur. Some of these transitions, the ones produced in the optical range, are forbidden by the parity selection rule ($\Delta L = \pm1$) and thus, the emitted lines are often called forbidden lines. Others, called semi-forbidden lines, only violate the spin rule ($\Delta S = 0$). 

The emission coefficient of a collisionally excited line produced by a radiative transition from level $k$ to level $l$ is given by: 
\begin{equation}
j_{kl} = \frac{h\nu_{kl}}{4\pi} f _{k}A_{kl}n(X^{+i}),
\end{equation}
where $f_k$ is the fraction of ions $X^{+i}$ in the upper level, $k$, and $A_{kl}$ is the spontaneous transition probability from level $k$ to $l$. To compute the emissivity of a CEL, one needs to know the population of the upper level. It is necessary to solve the statistical equilibrium equations: the rate of population of a level by radiative and collisional processes is balanced with the rate of de-population by these processes:
\begin{equation}
\sum_{l \neq k} f_{l}n_{e}q_{lk} + \sum_{l > k} f_{l}A_{kl} = \sum_{l \neq k} f_{k}n_{e}q_{kl} + \sum_{l < k} f_{k}A_{kl},
\end{equation}
where  $f_{l}$ and  $f_{k}$ are the fraction of ions $X^{+i}$ in the levels $l$ and $k$, and $q_{lk}$ and $q_{kl}$ are the collisional deexcitation and excitation rates. The detailed expressions of $q_{lk}$ and $q_{kl}$ can be found in the book by \citet{Osterbrock2006}. To solve these equations, atomic data (collision strengths and radiative transitions) and an estimate of the physical conditions are necessary. Softwares such as PyNeb \citep{Luridiana2015} and the {\sc iraf}\footnote{http://iraf.noao.edu/docs/spectra.html} package {\sc nebular} provide a solution for the equations for a number of levels depending on the ion. 

The CELs are one of the main cooling factors within ionized nebulae, so the presence of intense CELs depends on having available levels a few eV above the ground level. Since the temperature in photoionized nebulae is usually equivalent to about 1 eV, these differences are only available within the same ground state electron configuration, and therefore require the presence of a sub-shell with at least 2 electrons and 2 empty spaces to have the possibility of fine structure interaction between equivalent electrons to produce multiple levels within such configuration.

\subsubsection{Fluorescence}\label{sssec:elm_fluorescence}

Some permitted lines of some elements, such as oxygen, are brighter than expected by pure recombination because they are excited by starlight or by other nebular lines. The Bowen lines are one particular case of these lines, they are produced because there is a coincidence between the wavelengths of two lines; in this example, the \ion{O}{3} line at 303.80 \AA\ and the \ion{He}{2} line at 303.78 \AA. Some of the photons emitted by the He$^{++}$ atoms are absorbed by the O$^{++}$ atoms and then are reemitted. Since the interpretation of fluorescence lines is complicated, they are preferably not used to derive physical or chemical parameters in nebulae, but indeed, it is important to identify this type of lines in order to avoid them in the calculations. 

\paragraph{Case C and Case D}

From the early studies of photoionized regions it was recognized that the hydrogen spectrum strongly depends on the optical depth of the Lyman lines as well as on the effect of the posible presence of fluorescence of the same lines, \citet{Baker1938} and \citet{Aller1939} defined Case A as the simplest possible case, where the nebula is optically thin; Case B, when the size of the nebula is such that the optical depth of the Lyman lines is so large that the fraction of such photons that escape is negligible; and Case C when the spectrum of an optically thin nebula is affected by fluorescence of hydrogen due to non ionizing continuum \citep{Ferland1999}; Case B has been the most studied of these, since the physical conditions of most nebulae are closer to those of Case B than to those of cases A and C. Recently it has been recognized that some objects are affected by both, optical depth and fluorescence \citep{Luridiana2009, Peimbert2016} this scenario is frequently called Case D. 

Since, in Case D, the hydrogen transitions are optically thick, de-excitation occurs through higher series lines, in particular excitation to level $n$ $(n \geq 3)$ produce transitions to $n'$ $(n' \geq 2)$, therefore Balmer emissivities are systematically enhanced above case B predictions. Moreover the \ion{He}{1} lines are also enhanced by fluorescence. Case D produces small effects in the \ion{H}{1} and \ion{He}{1} lines but they might be important in the determination of the primordial helium abundance.

\paragraph{Optical depth of levels other than the ground level}The first excited level of H decays too fast to be of any significance; heavy elements are not abundant enough to have very large optical depths; however, He$^0$ is abundant enough, and has a metastable level with a long enough half-life to require special attention. 

The effects on the helium abundance determination due to the optical depth of the 2$^3$S metastable level have been studied by \citet{Robbins1968}, \citet{Benjamin2002}, and \citet{Aver2011}.

\section{Calculation of physical conditions and ionic abundances from observations} \label{sec:observations}

Using the concepts described in Section \ref{sec:processes}, and the simple assumptions that the photoionized region is homogeneous in temperature, density, and chemical composition, it is possible to determine the physical conditions and chemical abundances of photoionized regions. To a first approximation these three assumptions seem to be adequate, (all the gas has a common origin, and photoionization models show that temperature should vary by only a few percent across most of each photoionized region. Careful study of the best observed objects, in particular the presence of Abundance Discrepancy Factors (ADFs), show that at least one of those simplifications is not warranted.

In this section we will describe how, starting from observations of specific photoionized regions and atomic data sets, it is possible to determine the physical conditions (density and temperature) and the ionic chemical abundances. While presenting the determinations of ionic chemical abundances we will present two sets of chemical abundances: those derived from CELs and those derived from RLs; when both sets are available RLs produce higher chemical abundances, and their ratio is called the ADF.

\subsection{Quality of the available spectra}\label{ssec:quality}

Long-slit optical spectra have been widely used in the literature to derive the physical conditions and ionic abundances of ionized nebulae. There are spectra of several hundreds of Galactic and extragalactic PNe and \Hiiregs\ with resolutions better than $\sim5$ \AA\ \citep[see, for example, the compilations by][]{Kwitter2012, Maciel2017}. These spectra allow the detection of nebular lines such as [\ion{O}{3}] $\lambda$4959, $\lambda$5007, [\ion{N}{2}] $\lambda$6548, $\lambda$6583, [\ion{S}{2}] $\lambda$6717, $\lambda$6731, as well as RLs from hydrogen, helium, and carbon (\ion{C}{2} $\lambda$4267). Deep long-slit spectra also allow the measurement of weaker lines such as the auroral lines [\ion{O}{3}] $\lambda$4363 and [\ion{N}{2}] $\lambda$5755. Some available long-slit spectrographs are LRIS at Keck I telescope, GMOS at Gemini telescope, FORS at Very Large Telescope (VLT), and OSIRIS at Gran Telescopio Canarias (GTC); many more spectrographs are available in smaller telescopes.

Echelle spectra provide a resolution better than 1 \AA\ that allows us to separate and measure nearby lines, such as the mutiplet 1 of \ion{O}{2} at $\sim$4650, used to compute O$^{++}$ abundances with RLs, whose lines can be contaminated with lines of \ion{N}{2}, \ion{N}{3}, [\ion{Fe}{3}]. The number of ionized nebulae with deep and high resolution data in the literature is around 50 \citep[see, e.g. the sample used by][]{DelgadoInglada2014a}. Some examples of echelle spectrographs are HIRES at Keck I telescope, MIKE at Magellan Clay telescope, and UVES at VLT; dozens of other echelle spectrographs are available in smaller telescopes, but one requires echelles at large telescopes to obtain deep and high resolution spectra needed to study faint lines.

The use of Integral Field Unit (IFU) and multi-object spectrographs (MOS) allows the spatial study of ionized nebulae, often at the expense of a poorer spectral resolution. Several long-slit spectrographs are also MOS and some instruments may operate in both modes (MOS and IFU). A few examples of abundance studies based on IFU or MOS data can be found in \cite{Magrini2009, Stasinska2013, Kehrig2016, Zinchenko2016}. Some examples of MOS and IFU spectrographs are: GMOS at Gemini telescope, FORS  and VIMOS at VLT, and PMAS at the CAHA 3.5m telescope. With the arrival of instruments such as MEGARA \citep{GildePaz2016} in GTC, a high resolution IFU and MO spectrograph, it will be possible to study ionized nebulae with a high spectral and spatial resolution. 

\subsection{Calibrations}\label{ssec:calibrations}

Spectral emission lines allow the determination of physical conditions and ionic abundances in ionized nebulae. Before using spectroscopic data one should remove the effect of the instrument and the atmosphere and perform some calibrations. This is called data reduction and it depends on the type of spectra (long-slit, echelle, multi-fiber, integral field unit) and on the instrument and telescope. It is beyond the scope of this tutorial to explain in detail the whole process of data reduction and we refer the reader to the webpages of the observatories (such as GTC\footnote{http://www.gtc.iac.es/}, ESO\footnote{http://www.eso.org} and GEMINI\footnote{http://www.gemini.edu}) to find more information. One of the main packages used to reduce astronomical data is {\sc iraf}\footnote{http://iraf.noao.edu/docs/spectra.html} and there are many manuals available explaining how to use it. 

In brief, the main steps in data reduction are: bias and dark subtraction, cosmic rays removal, flatfield correction, wavelength and flux calibration. The bias is the number of counts in the detector pixels for zero second exposures. The dark current is the number of counts in the pixels when no light is falling in the detector and it is caused by the movements of electrons in the electronics. Both signals need to be subtracted from all images. The flatfield image allows the correction of pixel-to-pixel variations due to differences in the sensitivity. Cosmic rays are high-energy particles that arrive into the detectors constantly and need to be removed, which is easy when there are various exposures of the same field. Wavelength calibration consists in the transformation of pixel scale into wavelength scale and it is done by using comparison spectra of lamps (such as Th-Ar, Ne lamps). And finally, flux calibration transforms the number of counts into intensities and for this, standard stars are required. Besides, one would like to trim the spectra, combine them into one final spectrum, then correct from: the Earth movement around the Sun, the interstellar reddening, and the earth atmospheric lines if possible.

The observation of the continuum and the emission lines of gaseous nebulae need to be corrected for interstellar extinction due to interstellar dust. We refer the reader to the discussion of this issue in Chapter 7 of \citet{Osterbrock2006}.

\subsection{Continuum emission and underlying absorption in nebular spectra}\label{ssec:underlying}

The emission line spectra are superimposed on top of a weak continuum that has to be subtracted from the observations to recover the emission line intensities. The weak continuum is due to bound free and free-free emission, two photon emission, dust scattered light, and starlight. For galactic objects it is possible to avoid the central star in PNe and the brightest stars in galactic \Hiiregs, but not for extragalactic \Hiiregs\ and PNe. Star light will contain many permitted lines in absorption; of particular interest are the \ion{H}{1} lines since they are used to calibrate most of the objects and can be significantly affected by underlying absorption. The errors produced by the underlying component can be minimized by studying objects with the highest equivalent width of H$\beta$ in emission.

We consider that the best procedure to correct for underlying absorption in \Hiiregs\ is to use the models by \citet{GonzalezDelgado1999, GonzalezDelgado2005}. According to these models, for young objects the  EW$_{\rm ab}$(H$\beta$) is expected to be less than 2.5  \AA, and at the same time the EW$_{\rm em}$(H$\beta$) is expected to be more than $\sim150$; on the other hand, for older objects, EW$_{\rm ab}$(H$\beta$) is expected to be larger, while EW$_{\rm em}$(H$\beta$) is expected to be smaller. The correction for underlying absorption for objects with EW$_{\rm em}$(H$\beta$) $>$150 A is inversely proportional to EW$_{\rm em}$(H$\beta$), while for objects with EW$_{\rm em}$(H$\beta$) $\leq$ 150 \AA\ the correction, and consequently the associated error, increases even faster due to the larger EW$_{\rm ab}$(H$\beta$) predicted by the models.

\subsection{Atomic data}\label{ssec:det_data}

The first step, when calculating the physical conditions and ionic abundances is to select the set of atomic data for your calculations. A different selection will translate into a different result. A compilation of some of the most used atomic data was provided by \citet{Mendoza1983} and other recent compilations can be found on the CHIANTI \footnote{http://www.chiantidatabase.org/} and NIST \footnote{http://physics.nist.gov/} webpages. Atomic physicists make a great effort providing us accurate atomic data, it is fair to give them the credit by citing the original papers where the atomic data are published. 

\citet{JuandeDios2017} recently discuss the impact of different sets of atomic data on the determination of the chemical abundances of O, N, S, Ne, Cl, and Ar.
 
The package PyNeb contains several sources of atomic data so that the user can choose their preferred ones. There is also a default set of atomic data recommended by the developers of PyNeb. The software {\sc iraf}  does not allow a simple change of the atomic data, but the adopted sources can be checked. The C17 version of the photoionization code {\sc cloudy} \citep{Ferland2013} allows an easier treatment of the atomic data since the files have been moved to external databases.

\subsection{Determination of physical conditions}\label{sec:determination_physical}

The spectra of ionized nebulae show bright emission lines that allow us to determine the physical conditions of the gas: electron temperatures (\temp) and densities (\dens). 

The electron configurations with a complete subshell only have one specific configuration, and thus one available level; and ground configurations with a subshell with only one electron (as well as those lacking one electron to be complete), while theoretically able to have more than one configuration, have several configurations which are equivalent and thus have the same energy; for these 3 types of configurations the first excited state requires at least one electron to move from one electronic  shell to another. The amount of energy required for such transitions is: a) too high to be easily accesible via collisions, and b) lies in the ultraviolet. On the other hand ions with ground-state electronic configurations of $s^2p^2$, $s^2p^3$, and $s^2p^4$ are easily observed (which have their first four excited levels with an energy $\sim kT$, i.e., easily reachable via collisions and with some transitions in the optical range); the study of the intensities of these transition has long been understood and many codes are available which can model the intensities of such lines as a function of temperature, density, and abundance. In principle ions with electron configurations of the form: $s^2d^i$ ($2 \le i \le 8$), can also be studied in a similar manner, but ions with the required number of electrons are less abundant; also, since they have many more energy configurations available, software to model the behavior of such ions is less easy to obtain.

The $p^2$, $p^3$, and $p^4$ ions have line ratios sensitive to the electron temperature. The numbers of electrons required for those configurations are exactly 6, 7, 8, 14, 15, or 16; in principle ions with 32, 33, and 34 electrons have the same configurations, but the most abundant of these would be Se$^{++}$, which is 5 orders of magnitude less abundant than O$^+$ and is too faint for its auroral lines to be seen.  Some examples of observable ions are: N$^{+}$, O$^{+}$, O$^{++}$, Ne$^{++}$, S$^{+}$, S$^{++}$, Cl$^{++}$, Ar$^{++}$, and Ar$^{+3}$. 

The $p^3$ ions have line ratios sensitive to the \dens . The numbers of electrons required for this configurations are exactly 7 and 15; in principle ions with 33 electrons have the same configuration, but the most abundant of these would be Se$^{+}$, which is even less abundant than Se$^{++}$ (although the relevant lines for traditional density determinations are nebular lines). Some examples of observable ions are: O$^{+}$, Ne$^{+3}$, S$^{+}$, Cl$^{++}$, and Ar$^{+3}$.

\begin{figure}[ht!]
\centering
\includegraphics[trim = 60 0 0 0, clip, width = 9cm]{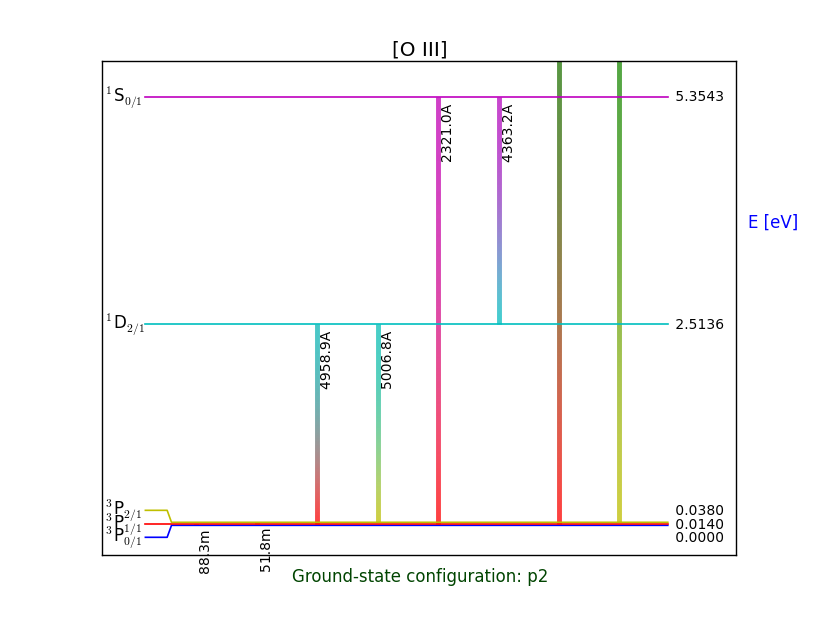}
\caption{Grotrian diagram for O$^{++}$ from PyNeb \citep{Luridiana2015}\label{fig:O3}}
\end{figure}

\begin{figure}[ht!]
\centering
\includegraphics[trim = 60 0 0 0, clip, width = 9cm]{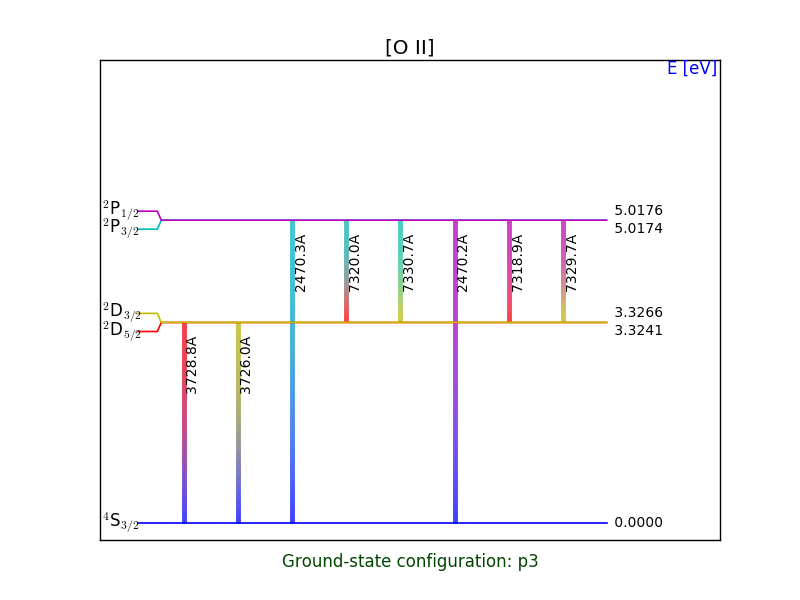}
\caption{Grotrian diagram for O$^{+}$ from PyNeb \citep{Luridiana2015}.\label{fig:O2}}
\end{figure}

\begin{figure}[ht!]
\centering
\includegraphics[trim = 60 0 0 0, clip, width = 9cm]{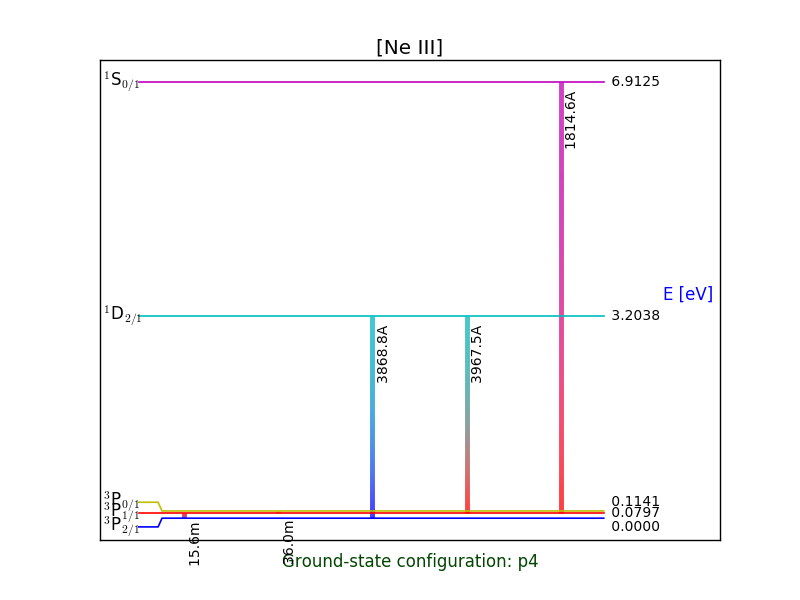}
\caption{Grotrian diagram for Ne$^{++}$ from PyNeb \citep{Luridiana2015}.\label{fig:Ne3}}
\end{figure}

The Grotrian diagrams for O$^{+}$, O$^{++}$, and Ne$^{++}$ obtained with PyNeb are presented in Figures~\ref{fig:O3}--\ref{fig:Ne3}.

In principle we can also use $d^i$ levels ($2 \le i \le 8$) to calculate physical conditions, but  the number of excited energy levels that need to be considered, as well as the atomic physical parameters makes it impractical, and only a few of these ions have been studied (e.g. Fe$^{++}$, Fe$^{+3}$, and Ni$^{++}$)  and numerical packages used to study them are not widely distributed (PyNeb does contain these ions).

\subsubsection{Electron temperatures from CELs}\label{ssec:temp_cel}

The intensity ratios of some forbidden lines are highly sensitive to \temp\ and thus, are very useful to calculate it. The reason is that electrons with very different energy are needed in order to populate the different ionic levels of an ion through collisions. 

For example, the excitation energy of level $^1{\rm S}_0$, from which the [\ion{O}{3}] $\lambda4363$ line originates, is 5.35 eV whereas the excitation energy of level $^1{\rm D}$, from which the [\ion{O}{3}] $\lambda5007$ line originates, is 2.51 eV (see Fig. \ref{fig:O3}). Therefore, the ratio between these two lines tells us about the temperature of the plasma. We expect the intensity ratio of [\ion{O}{3}] $\lambda4363/\lambda5007$ to be higher in hotter nebulae. 

In Chapter 5 of the book by \citet{Osterbrock2006} the reader can find the analytical expressions for the intensity ratio [\ion{O}{3}] $(I(4959)+I(5007))/I(4363)$ and other intensity ratios that are commonly used in the literature to estimate $T_e$,  such as [\ion{N}{2}] $I(5755)/(I(6548)+I(6583))$, [\ion{Ne}{3}] $I(3343)/(I(3869)+I(3968))$, and [\ion{S}{3}] $I(6312)/(I(9532)+I(9069))$. 

The main disadvantage is that the auroral lines (such as [\ion{O}{3}] $\lambda4363$ and [\ion{N}{2}] $\lambda5755$) are weak and, if the observations are not deep enough, they are not detected. Alternative methods have been proposed to compute the chemical abundances when the \temp\ cannot be derived, such as the so-called strong line methods that will be discussed in Section \ref{ssec:total_strong_line}.

\subsubsection{Electron temperatures from RLs}\label{ssec:temp_rl}

The intensity of a recombination line is approximately proportional to $T^{-1}$, this approximation would result in ratios that are completely independent of the temperature; yet, when studied in detail, it is found that $I_{rl}\propto T^{-1} \Upsilon$, where $\Upsilon$ is the oscillator strength of each transition, and has a weak temperature dependance which locally can be represented as  $\Upsilon \propto T^{\kappa}$ where $-0.2 \lesssim \kappa \lesssim 0.2$, this approximation shows a small dependance on the temperature that can, occasionally, be exploited to determine the temperature using only RLs. 

In reality this is very difficult, since the hydrogen   lines have a nearly homogeneous dependence on $T$, and RLs of heavy elements are too weak to be used. The only remaining possibilities are temperatures determined from He, where the errors are at least $\sim1000$K for the best observed objects \citep[e.g.][]{Peimbert2000,Peimbert2007,Izotov2007}, or to try to combine one RL with a CEL \citep{Peimbert2013}; in this last scenario, although the sensitivity of the ratios is similar to the sensitivity of nebular to auroral CEL ratios, the intensity of RLs is frequently smaller than that of auroral lines, thus the observational errors tend to be large.

\subsubsection{Electron temperatures from the Balmer continuum}\label{ssec:temp_balmer}

In a gaseous nebula, it is possible to determine the temperature from the ratio of the Balmer jump to a Balmer line, this because the intensity of any Balmer line as well as the total energy in the Balmer continuum is proportional to $n$(H)$\times T^{-1}$, however the Balmer continuum becomes wider (in wavelength range) with increasing temperature ($\propto T^{1/2}$), therefore the height of the Balmer discontinuity is approximatelly proportional to $T^{-1.5}$.

To determine the temperature from the Balmer continuum it is necessary to estimate the underlying Balmer discontinuity in absorption due to the direct star contribution and to the dust scattered light. The effect of the underlying Balmer discontinuity due to the stellar continuum on the determination of the temperature of the nebulae becomes negligible for a radiation field dominated by stars hotter than about 45\,000K (i.e. it is important for all \Hiiregs, but can be ignored for most PNe).

Balmer temperatures have been determined by several groups \citep[e.g.][]{Peimbert1969,Liu1993}.

\subsubsection{Temperature inhomogeneities ($t^2$)}\label{ssec:t2}

As we said before, photoionized nebula are frequently assumed to have uniform temperature; yet photoionization codes already show some temperature variations across each nebula, and the existence of other physical processes suggests the existence of larger variations. 

To a second order approximation we can characterize the temperature structure of a gaseous nebula by two parameters: the average temperature, $T_0$, and the root mean square temperature fluctuation, $t$, given by:
\begin{equation}
\label{equ:T0}
T_0(X^{+i})=
{{\int T_e n_e n(X^{+i})\,dV}
\over{\int n_e n(X^{+i})\,dV}},
\end{equation}
and
\begin{equation}
\label{equ:t2}
t^2(X^{+i})=
{\int {\left(T_e - T_0(X^{+i}) \right)^2 n_e n(X^{+i})\,dV}
\over{T_0(X^{+i})^2 \int n_e n(X^{+i})\,dV}},
\end{equation}
respectively, where $n_e$ and $n(X^{+i})$ are the electron and the ion densities of the observed emission line and $V$ is the observed volume \citet{Peimbert1967}.

To determine $T_0$ and  $t^2$  we need two different measurements of $T_e$: one that weights preferentially the high temperature regions and one that weights preferentially the low temperature regions of the observed volume. For example the temperature derived from the ratio of the [\ion{O}{3}] $\lambda\lambda$ 4363, 5007 lines, $T_e(4363/5007)$, and the temperature derived from the ratio of the Balmer continuum to $I$(H$\beta$), $T_e({\rm Bac/H}\beta$); these temperatures are related to $T_0$ and $t^2$ by:
\begin{equation}
\label{equ:Te(4363/5007)}
T_e(4363/5007)=
T_0({\rm O ^{++}})
\left[ 1 + {1 \over 2} \left( {91300 \over T_0({\rm O ^{++}})} - 2.68\right) t^2 \right],
\end{equation}
and
\begin{equation}
\label{equ:Te(Bac)}
T_e({\rm Bac}/{\rm H}\beta)=
T_0({\rm H ^{+}})
\left( 1 - 1.70 t^2 \right),
\end{equation}
respectively \citep{Peimbert2014}. These two equations are very good approximations to $T_0$ and $t^2$, 
when terms of higher order in $t$ can be neglected, that is when $t^2 \ll 1.00$. 

It is also possible to determine a temperature from the intensity ratio of a collisionally excited line of an element $p + 1$ times ionized to a recombination line of the same element $p$ times ionized, for example the ratio of [\ion{O}{3}] $\lambda$5007 to the RLs of multiplet 1 of \ion{O}{2}, $T_e$(\ion{O}{2}/[\ion{O}{3}]), this ratio is independent of the element abundance and depends only on the electron temperature. By combining  $T_e$(\ion{O}{2}/[\ion{O}{3}]) with a temperature determined from the ratio of two CELs, like $T_e(4363/5007)$, it is also possible to derive $T_0$ and $t^2$ \citep[e.g.][]{PenaGuerrero2012b,Peimbert2013}.

Another temperature, $T_e$ (\ion{He}{1}), can be obtained from the intensity of many pairs of \ion{He}{1} RLs, because each line has a slightly different temperature dependance \citep[e.g.][]{Peimbert2000, Izotov2007, Porter2013}.

Most of the \Hiiregs\ observed in other galaxies are very bright and have most of its oxygen in the O$^{++}$ stage and most of its helium in the He$^+$ stage. Therefore the  $t^2$ values derived  from $T_e(4363/5007)$  together with any of the following temperatures: $T_e({\rm Bac}/{\rm H}\beta)$,   $T_e$(\ion{O}{2}/[\ion{O}{3}]), or $T_e$ (\ion{He}{1}) are representative of the whole object.

According to its definition, $t^2$ can be determined even when the material is not chemically homogeneous (as is expected in some PNe). However, under such circumstances, the $t^2$ of each element can be completely different; moreover, when any element is not well mixed with hydrogen, it is not possible to define observationally the total abundances.

The net effect of $t^2$ on the temperatures is that temperatures derived from CELs are larger than $T_0$, which in turn is larger than temperatures derived from RLs (temperatures derived from the ratio of a CEL to a RL tend to be similar to $T_0$). For CELs this effect is larger at small $T_e$, while for RLs this effect is larger for large $T_e$.

\subsubsection{Local electron densities from CELs}\label{sssec:dens_forb}

The \dens\ can be derived from the intensity ratio of CELs of the same ion originated from levels with nearly the same excitation energy. Thus, the intensity ratio does not depend on \temp\ but does depend on the ratio of the collision strengths. If the involved lines have different transition probabilities or different collisional deexcitation rates, their intensity ratio will strongly depend on \dens.  

Some of the intensity ratios used to derive \dens\ are: [\ion{O}{2}] $\lambda3726/\lambda3729$, [\ion{S}{2}] $\lambda6717/\lambda6731$, [\ion{Cl}{3}] $\lambda5518/\lambda5538$, and [\ion{Ar}{4}] $\lambda4711/\lambda4740$. Each intensity ratio is valid in a specific range of densities. In the low density regime the intensity ratio is proportional to the collisionally excitation ratio whereas in the high-density regimen (\dens\ above the critical density), collisions dominate and the intensity ratio is proportional to the spontaneous transition probability ratio. Chapter 5 of the book by \citet{Osterbrock2006} explains further details about the determination of \dens. 

In general, \temp\ and \dens\ from forbidden lines are computed together since they depend on each other. Softwares such as PyNeb \citep{Luridiana2015}, Neat \citep{Wesson2012}, and the routine {\sc temden} from {\sc iraf} \citep{Shaw1995} calculate the physical conditions from emission line spectra. 

It is important to mention that \temp\ and \dens\ computed from one particular intensity ratio are representative of the zone of the nebula where the involved intensity lines are emitted. When calculating ionic abundances one has to make assumptions on the temperature and density structure of the nebula. For example, if \temp\ and \dens\ are homogenous throughout the entire nebula or if there are different regions in the nebula with different representative physical conditions.

\subsubsection{Local electron densities from RLs}

There are at least two ways to determine densities from RLs: the metastable $2\,^3S$ level of \ion{He}{1}, or the density dependence of the lines of the mutliplet 1 of \ion{O}{2}. Due to the specific characteristics of He, no other ion, with the same characteristics, is expected to be abundant enough to allow us to use the same technique; on the other hand, other \ion{O}{2} multiplets, as well as multiplets of many other ions should have similar density dependences as the multiplet 1 of \ion{O}{2}, but, since they will be fainter, none of them has been studied so far.

When He$^+$ recombines, the new electron can be either parallel or anti-parallel to the old one, creating 2 families of excited He$^0$, while the anti-parallel family will quickly reach the ground state ($1\,^1S$), the helium  that recombine with parallel electrons will arrive at the metastable $2\,^3S$ level and can easily remain there for a fraction of an hour. The energy structure is such that collisions with helium  in the ground state cannot excite the atom, while collisions with helium in the metastable level can; this will affect each \ion{He}{1} line differently and its efficiency depends on the density, which makes it possible to use this effect to determine the density. Detailed calculations of the atomic physics of He$^+$ recombination can be found in \citet{Porter2013}.

The line ratios of the multiplet 1 of \ion{O}{2} depend strongly on $n_e$ \citep{Peimbert2005a}; this dependence arises in the O$^{++}$ before its recombination. At high densities O$^{++}$ is expected to be distributed in a 3:2:1 ratio between its three lowest energy states, while at low densities all the ions are expected to be at the ground energy level; even after recombination this signatures are not completely erased \citep{Bastin2006}. The expected ratios for the lines of multiplet 1 of \ion{O}{2} as a function of $n_e$ can be found in \citet{Peimbert2005b} and  \citet{Storey2017}. 

\subsubsection{Root mean square densities and filling factor}\label{sssec:dens_rms}

In their pioneering study of \dens\ in nebulae, \citet{Seaton1957}, noted that density inhomogeneities produced a disagreement between densities derived from forbidden line ratios and those derived from H($\beta$) surface brightness; an example quoted by them was NGC 7027. \citet{Osterbrock1959} studied in detail the density distribution in the Orion nebula and showed that the densities derived from radio fluxes were considerably smaller than those derived from the [\ion{O}{2}] 3726/3729 line intensity ratio, they suggested a model in which only a fraction $\epsilon = 0.03$ of the nebula was filled with high density material and the rest was empty. Since then $\epsilon$ has been called the filling factor and is defined by
\begin{equation}
\epsilon = \frac{n_e^{2}(\rm{rms})}{n_e^{2}(\rm{CEL})},
\end{equation}
where, \dens(FL) is the density determined from forbidden line ratios and \dens(rms) is the root mean square density determined from a Balmer line or from the radio continuum flux.

In the presence of a filling factor the mass of a nebula is given by:
\begin{equation}
M(\epsilon) = \epsilon^{1/2} M(\rm{rms}).
\end{equation}                          

For various epsilon determinations of PNe see \citet{TorresPeimbert1977} and \citet{Mallik1988} and references therein.

\subsubsection{Other density distributions}\label{sssec:dens_other}

The filling factor is only an approximation, real objects will not have $\sim97\%$ of their volume with perfect vacuum while the other $\sim3\%$ has a constant density; such a model can only be considered a first approximation to reality. With the wide spread presence of ADFs models within this approximation seem incapable of reproducing the observations. As long as the densities are below the critical density for all CELs, and the material is well mixed, reality can be approximated by a filling factor plus constant density.

However, if there are regions with density above the critical density for nebular lines, the intensity of such lines will no longer be proportional to density squared; to ignore this will cause us to measure an unexpectedly low intensity nebular line, which will make us a) underestimate the abundance of such ion, b) if used to determine the temperature, to overestimate the temperature, and c) when using such high temperature, to underestimate the abundance (again). Also, when enough nebular and infrared lines, of different ions, get suppressed the cooling will get affected, and a temperature structure will emerge.

If the gas is not well mixed and there is a density structure correlated with the chemical composition traditional line ratio analyses no longer make sense; instead models of the expected line intensities have to be made. Many ``2-phase" models are explored in the literature \citep[e.g.][]{Ercolano2003, Yuan2011}; in general these models include a semi-large hot ``normal" phase, small droplets of a cold ``hydrogen-poor" phase, and a large vacuum volume. Of course, these phases are required when neither phase is close enough to reproduce the observations, so the phases will be very different and the average (sum) of their emissions will be very different from the emission of their of a phase with average physical conditions; yet it is unlikely that, in real objects, all the droplets will have the exact same density and chemical composition, also there is bound to be a transition layer between the droplets and the surrounding medium (as well as between the hot phase and the vacuum); so, when required, these 2-phase models should be considered only as an approximation to reality.

\subsection{Determination of ionic abundances}\label{sec:ionic}

As in most other ISM environments, in photoionized regions, hydrogen is the most abundant element, representing $\sim 90\%$ of the available atoms; these regions are considered ionized if most of the atoms are ionized; therefore, in most photoionized regions, ${\rm H}^+$ is the most abundant ion comprising $\sim 90\%$ of the available ions. For this reason, most of the time, abundances are measured relative to $n$({\rm H}$^+$), as $n(X^{+i})/n({\rm H}^+)$ or as $12+\log(n(X^{+i})/n({\rm H}^+))$. For simplicity from now on, we will use the following notation: $X^{+i}/{\rm H}^+$ and $12+\log(X^{+i}/{\rm H}^+)$.

A side effect of this definition is that there are no pure CEL abundance determinations, since all abundances are made with respect to H$\beta$ (or some other hydrogen line).

There is an additional subtlety regarding ionic abundance determinations: when determining abundances, we make sure to use the best available density for $X^{+i}$ (the ion we are trying to determine), but we make use of the same density to represent  ${\rm H}^+$, when we know that, most of the time, the representative abundance for ${\rm H}^+$ can be substantially different; we will even use a different density for ${\rm H}^+$ when determining the abundance for a different ion. A simple example of this can be studied by considering an object with ${\rm O/H}=0.001$ where 50\% of the oxygen is singly ionized, and the other 50\% is twice ionized; let us further assume that the representative density for ${\rm O}^+$ is  $n_e({\rm O}^+)= 100 \, {\rm cm}^{-3}$ and for ${\rm O}^{++}$ is  $n_e({\rm O}^{++})= 400 \, {\rm cm}^{-3}$. Under these assumption ${\rm O}^+$ occupies 80\% of the volume , while  ${\rm O}^{++}$ occupies the other 20\%; on the other hand, if we weigh by emission measure, we find that the  ${\rm O}^+$ region produces only 20\% of the intensity of H$\beta$, while the ${\rm O}^{++}$ region produces the other 80\%. In this example, to obtain ${\rm O}^{+}/{\rm H}^+={\rm O}^{++}/{\rm H}^+= 0.0005$, the density we should use for  ${\rm H}^+$ is:  $n_e({\rm H}^+)= 200 \, {\rm cm}^{-3}$. Instead, to determine ${\rm O}^{+}/{\rm H}^+$, we assume $n_e({\rm O}^+)= n_e({\rm H}^+)= 100 \, {\rm cm}^{-3}$, when doing this we find ${\rm O}^{+}/{\rm H}^+= 0.0002$ (since, at such density, a lot more ${\rm H}^+$ is needed to produce the intensities observed in the high density region); at the same time, to determine ${\rm O}^{++}/{\rm H}^+$, we assume $n_e({\rm O}^{++})= n_e({\rm H}^+)= 400 \, {\rm cm}^{-3}$, when doing this we find ${\rm O}^{++}/{\rm H}^+= 0.0008$ (now a lot less ${\rm H}^+$ is needed to produce the intensities observed in the low density region). Although the final result, when determining total abundances, is correct,  ${\rm O}/{\rm H}= 0.001$; one should be careful of the true meaning of measuring ${\rm O}^{++}/{\rm O}^+=4.0$.

What the previous example shows, is that we implicitly weigh our observations by emission measure; not only is this the most natural way of weighing observations, it has additional advantages since most physical proceses in photoionized regions are proportional to density squared (e.g. emission, recombination, absorption, the rate at which ionizing photons are consumed, collisions), and, at the end of the day, if the material is well mixed, it gives us the correct total abundance; however there are still many processes that do not depend on the density squared (e.g. the degree of ionization partially depends in density, computer models cannot ignore density), in particular, in the presence of chemical inhomogeneities, we are not able to reproduce the total abundance with such ease, instead detailed (complex) models are required.

\subsubsection{From RLs}\label{ssec:ionic_recombination}

Although, to some extent, the terms ``permitted lines" and ``recombination lines" are often used interchangeably, one must take care not to confuse fluorescent lines (which will tend to be permitted lines) with RLs.

To a first approximation the intensities of most RLs are proportional to $n_e \times n(X^{+i})$ and inversely proportional to $T_e$; this remains so through more than an order of magnitude in $T_e$ as well as for densities of $10^{10}\, {\rm cm}^{-3}$, or even more (although there are a few important departures from such proportionality); this is because the number of interactions between ions and electrons is proportional to the density of both, because the electron capture efficiency increases for slower electrons, and because the emission occurs very fast, so very high densities are required for the ions to be collsionally de-excited. Also, since they are proportional to the ionic abundance, the only intense RLs are the lines originating in the most abundant ions. Of them the most important ones are the \ion{H}{1} lines, without which it is not possible to determine abundances \citep[the expected emissivities of \ion{H}{1} lines can be found in][]{Storey1995}; for most objects the atomic data to be utilized is the one for case B.

Other than hydrogen, helium lines are the RLs most utilized for abundance determinations, since helium is about 100 times more abundant than the third most abundant element; also, for temperatures below 50000K it is very inefficient to collisionally excite either ${\rm He}^0$ or ${\rm He}^+$. ${\rm He}^0$ has many more lines than ${\rm H}^0$, since it does not have the degeneracy on the energy levels present (e.g. there are 6 \ion{He}{1} lines that represent transitions of electrons from quantum number $n=4$ to $n=2$, while, for \ion{H}{1}, there is only 1 such line); however, one must be careful when choosing among these lines and select lines that are relatively intense, but not affected by collisions (e.g. \ion{He}{1} $\lambda\lambda$ 6678 or 4921). The presence of ${\rm He}^{++}$ produces a different set of problems, since many \ion{He}{2} lines can be blended with the \ion{H}{1} lines, and one must be sure the intensity one measures at 4861 \AA\ truly corresponds to H$\beta$ (the intensity of  \ion{He}{2} $\lambda$5412 is a good indicator of contamination by \ion{He}{2} lines to H$\alpha$ and H$\beta$); the abundance of ${\rm He}^{++}$ is best determined from \ion{He}{2} $\lambda$4686, which is the strongest \ion{He}{2} optical line.

Recombination lines of heavier elements are much fainter, and thus there are few spectra with enough signal to noise to determine heavy element abundance accurately. Some of the most famous are the \ion{C}{2} doublet $\lambda$4267 and the \ion{O}{2} octuplet $\lambda$4650; \ion{C}{2} $\lambda$4267 has been important for many decades now, not only is the brightest of the optical RLs from heavy elements, but since no carbon CELs are available in the visible part of the spectrum \ion{C}{2} $\lambda$4267 is the only way to determine carbon abundance; on the other hand the \ion{O}{2} $\lambda$4650 multiplet has been important since its lines are the brightest RLs from an ion with available CEL data, and as such, they represent the best way to measure an ADF. But, for a few objects, recombination lines of more than two dozen ions have been observed.

One big advantage of determining abundances from RLs, is that all have very similar dependences with $T_e$ and $n_e$; as such any error in the measured physical parameters, as well as the presence of a large $t^2$, does not affect the determinations. The main disadvantage is that, unless one is observing a relatively bright gaseous nebula with very large telescopes, there are only a handful of ions for which RLs are available (of course, in the presence of chemical inhomogeneities, RL determinations become meaningless, but CEL determinations become meaningless too).

\subsubsection{From CELs}\label{ssec:ionic_collisional}

Most of the emission lines emitted by ionized nebulae are optically thin and thus, their intensities are proportional to the abundance of the ion that emits the line. The intensity of CELs highly depends on \temp\ and therefore, a reliable estimation of the physical conditions is needed to obtain reliable ionic abundances. The general expression of an emission line has been presented in Eq.~\ref{equ:emiss}. The ratio between a CEL and H$\beta$ is:
\begin{equation}
\frac{I(\lambda)}{I({\rm H}\beta)}= \frac{X^{+i}}{{\rm H}^+}\frac{\nu_\lambda}{\nu_{{\rm H}\beta}}\frac{A_{nn'}f_k}{n_e \alpha^{eff}_{{\rm H}\beta}},
\end{equation}
where $f_k$ is the fraction of ions $X^{+i}$ in the upper level, $k$. Then, the ionic abundance can be expressed as:
\begin{equation}
\frac{X^{+i}}{{\rm H}^+} = \frac{I(\lambda)}{I({\rm H}\beta)} \frac{\epsilon_{{\rm H}\beta}}{\epsilon_\lambda},
\end{equation}
where the emissivities are derived by solving the statistical equilibrium equations to obtain the level populations. 

When deriving ionic abundances one has to adopt the adequate \temp\ and \dens\ for each ionic species. One simple assumption is that the whole nebula can be described with one \temp\ and one \dens. One can also calculate each ionic abundance with different \temp\ and \dens\ values. The abundances derived with each approach may be very different (not to mention the differences associated to the adopted atomic dataset).

\subsection{Abundance discrepancy factors (ADFs)}\label{ssec:ionic_adf}

In many objects it is possible to measure abundances of the same ion, from both RLs and CELs; they never agree (unless the error bars are very large). It is systematically found that abundances derived from RLs are higher than those derived from CELs. The ADF was defined by \citet{Tsamis2003} as:
\begin{equation}
ADF(X^{+i})=
{{X^{+i}_{\rm RLs}}\over{X^{+i}_{\rm CELs}}},
\end{equation}
and it is overwhelmingly determined to be greater than one. 

A note of warning: since abundances are usually presented in a logarithmic scale, in some works the ADF is defined also in a logarithmic scale where:
\begin{equation}
ADF^* = \log(ADF(X^{+i})) = \log\left( \frac{X^{+i}_{\rm RLs}}{{\rm H}^+}\right)  - \log\left( \frac{X^{+i}_{\rm CELs}}{{\rm H}^+}\right);
\end{equation}
when this convention is used the values are usually in the $0.2<ADF^*<0.8$ range. We recommend the use of the definition of Eq.~(17). 

Typical \Hiiregs\ present $1.5\lesssim ADF \lesssim 3$ while typical PNe present $2\lesssim ADF \lesssim 5$. Chemically inhomogeneous PNe show ADF values in the 10--80 range, with some PNe showing knots, in their inner riegions, with ADF values reaching as high as 800 \citep{Corradi2015}.

The presence of ADFs (different than 1.0) show that the simplest models, where the chemistry, temperature, and density are homogeneous (or even models where the chemistry is homogeneous, and there are 2 zones defined by their ionization degree), are not adequate to represent real photoionizated regions. 

Another important point about ADFs: in all objects where CEL and RL abundances can be simultaneously determined, ADFs have proven to be ubiquitous, and until their origin is not well understood all CEL abundances are suspect (for most objects RL abundances are, probably, a good approximation to the true abundances; yet, probably, for very few objects CEL abundances are a good approximation, see section \ref{sec:t2_large}). Thus, for objects where only CEL abundances are available, these values should be considered a lower limit to the true abundances; therefore a correction should be made \citep[e.g][]{PenaGuerrero2012a}; moreover, corrections available at present are only crude corrections, and much work should be done before having high quality corrections at our disposal.

In section \ref{sec:t2_large} we will discuss the physical processes which may be responsible for the observed ADFs.

\section{Calculation of total abundances}\label{sec:total}

There are different approaches to compute chemical abundances of PNe and \Hiiregs. One may compute the total abundances by adding up all the ionic abundances of each element. This can be done with RLs or CELs. Since there are significant differences between both ionic abundances, it is important not to mix them in the derivation of total abundances. When some of the ions are not observed, ionization correction factors (ICFs) need to be used. This is the more direct method to compute chemical abundances in ionized nebulae. As we mentioned above, the abundances from CELs and RLs are different, causing an ADF. One may use the $t^2$ formalism to solve the discrepancy. One may also compute total abundances without calculating first the ionic abundances. One option is fitting a photoionization model to the observations. Another is to use the so-called strong line methods. The following subsections explain in detail each of the aforementioned methods.

\subsection{Adding ionic abundances}\label{sssec:total_adding} 

The total abundance of one particular element is given by the sum of the ionic abundances of all the ions present in a nebula. If we can measure emission lines of all these ions, then the calculation of the total abundance is straightforward:
\begin{equation}
\frac{X}{{\rm H}}=\Sigma_i\frac{X^{+i}}{{\rm H}^+}. 
\end{equation}
In general, the available observations cover only a particular wavelength range (optical, ultraviolet, or infrared) or some of the lines are too faint to be measured and thus, the total abundances must be calculated only from a few ions. In this case, ionization correction factors (ICFs) must be used to take into account the contribution of those ions for which we cannot derive their ionic abundance:
\begin{equation}
\frac{X}{{\rm H}}=\Sigma_{{\rm obs}}\frac{X^{+i}}{{\rm H}^+} \times ICF. 
\end{equation}

The first ICFs were defined according to similarities between ionization potentials (IP) of different ions (e.g., \citealt{Peimbert1969}, \citealt{Peimbert1971}). For example, the widely used relations N/O = N$^{+}$/O$^{+}$ and Ne/O = Ne$^{++}$/O$^{++}$ are based on the similarities between the IP of N$^{+}$ and O$^{+}$ (29.6 eV and 35 eV, respectively) and Ne$^{++}$ and O$^{++}$ (63.4 eV and 54.9 eV, respectively). However, these simple relations have proved to be inadequate because do not take into account all the physics involved in the ionized gas (for example, charge exchange reactions). 

ICFs derived from photoionization models are more reliable because, in principle, they take into account all the physics involved in photoionization. The first important compilation of ICFs based on photoionization models (about a dozen models) is the one by \cite{Kingsburgh1994} but some of them have been improved considerably based on nets of more complex photoionization models. New ICFs derived in the last years from large grids of photoionization models are mentioned in Section \ref{sssec:ICF_PNe}.

Softwares such as PyNeb \citep{Luridiana2015}, {\sc neat} \citep{Wesson2012}, and the routine {\sc temden} from {\sc iraf} \citep{Shaw1995} calculate the physical conditions and ionic abundances from emission line spectra. In addition, PyNeb contains several ICFs that can be used to determine total element abundances.

\subsection{Fitting photoionization models}\label{ssec:total_models}

In some cases there are no available ICFs to compute the total abundances (e.g., for fluorine, phosphorous, and germanium). In others, the available ICFs are not valid. For example, some ICFs are not adequate for objects with very low or very high degree of ionization. In these situations, the chemical composition of ionized nebulae can only be obtained by constructing a tailored photoionization model. 

Photoionization models are the theoretical representations of real nebulae. They are computed with specific codes that include all the physics involved in photoionized models, they solve the ionization and energy equilibrium equations and calculate the radiation transfer. The most popular and used photoionization code is {\sc cloudy} developed by a large group of people led by G. Ferland \citep{Ferland2013}. The library PyCloudy \citep{Morisset2014a} can be used to compute pseudo-3D {\sc cloudy} models. Other commonly used photoionization codes are: {\sc mappings} \citep{Dopita2013} and {\sc mocassin} \citep{Ercolano2003}. 

A detailed recent review on photoionization models of PNe has been recently published by \citet{Morisset2016}. A previous paper on this topic is that by \citet{Ferland2003}.

To compute a photoionization model one should provide some information about the ionizing source and the ionized gas: the shape and intensity of the radiation, the chemical composition of the gas and dust in the nebula, and the geometry of the cloud. The output consists of the several physical quantities (such as the ionization structure and electron temperature at every position in the nebula) and the line intensities. The predicted intensities are then compared with the observations. In principle, a good agreement indicates that the input parameters are consistent with the observations. If the agreement is not reasonable, the input parameters need to be changed until observations and predictions agree. 

It is important to mention that high quality spectra are needed to obtain a good photoionization model. Adjusting only a few observables (even if those are all you have) should not give you confidence in the model you are creating. The more constraints you have, the better your computed photoionization model will be. Along with the emission lines, it is crucial to have an idea of the morphology of a given nebula and its density structure. Some examples of tailored models constructed in the last years are: IC 418 by \citet{Morisset2009}, TS01 by \citet{Stasinska2010}, and NGC~6302 by \citet{Wright2011}.

One sticking point from such models is that, when observations of RLs and CELs originating from the same ion are available, they are unable to reproduce the observed ADFs. Yet, for objects where the explanation for the observed ADF is a chemically homogeneous $t^2$, photoinoization models stil have advantages over the direct method: since the intensity of [\ion{O}{3}] $\lambda$5007 is less affected by $t^2$ than the [\ion{O}{3}] 5007/4363 ratio, models that adjust nebular lines will be less afected than calculations that use auroral to nebular ratios to determine abundances.

\subsubsection{Databases}\label{sssec:fitting_databases}

In addition to tailored models one may compute grids of photoionization models covering a wide range of physical parameters to study general behaviors of ionized nebulae. 

The Mexican Million Models database (3MdB, \citealt{Morisset2015}) contains several grids of models computed with {\sc cloudy}: more than half a million of photoionization models for PNe \citep{DelgadoInglada2014a}, $\sim40000$ models of diffuse ionized gas \citep{FloresFajardo2011}, $\sim85000$ models based on CALIFA observations \citep{Morisset2016}, and $\sim30000$ models of giant \Hiiregs\ that have been used to derive oxygen and nitrogen abundances \citep{ValeAsari2016}. All these models are available for the community and can be used to explore many open issues.

\subsection{Strong lines methods}\label{ssec:total_strong_line}

The last of the methods relies only in measuring the intensities of nebular lines, frequently normalized to H$\beta$. These methods are not designed to determine abundances of individual elements; but rather, by assuming that there is a proportionality in the abundances of all heavy elements, they are designed to determine the overall abundance of heavy elements. These methods look for lines, or line combinations, whose overall intensity depends clearly in the total metallicity, but only marginally in other aspects such as degree of ionization or density.

The most famous of the strong line methods was proposed by \citet{Pagel1979}, and is commonly known as Pagel's method or $R_{23}$, where 
\begin{equation}
R_{23}={I(5007)+I(4959)+I(3727) \over I(H\beta)}.
\end{equation}
One particular problem of $R_{23}$ is that it is bivalued, with $R_{23}$ reaching a maximum of $\sim10$ when $Z\sim1/3Z_\odot$, and decreasing for both larger and smaller metallicities; another problem, common to most methods, is that it has a (weak) dependence on ionization degree.

Besides $R_{23}$ there are at least a dozen other strong line methods such as: [\ion{O}{3}]/H$\beta$ \citep{Aller1942}, [\ion{N}{2}]/H$\alpha$ \citep{StorchiBergmann1994}, [\ion{O}{2}]/H$\alpha$ \citep{Thompson1991}, [\ion{N}{2}]/[\ion{O}{2}] and [\ion{N}{2}]/[\ion{S}{2}] \citep{Jensen1976}, [\ion{O}{3}]/[\ion{N}{2}] \citep{Alloin1979}, [\ion{S}{2}]/H$\alpha$ \citep{Denicolo2002}, ([\ion{S}{2}]+[\ion{S}{3}])/H$\alpha$ \citep{Vilchez1996}, ([\ion{S}{2}]+[\ion{S}{3}]+[\ion{S}{4}])/H$\alpha$ \citep{Oey2000}, ([\ion{O}{2}]+[\ion{Ne}{3}])/H$\gamma$ \citep{PerezMontero2007}, [\ion{Ne}{3}]/[\ion{O}{2}] \citep{Nagao2006}, and [\ion{Ar}{3}]/[\ion{O}{3}] and [\ion{S}{3}]/[\ion{O}{3}] \citep{Stasinska2006}. Many of them were selected to avoid the bivaluation of $R_{23}$, or to avoid the necessity of the very blue [\ion{O}{2}] $\lambda$3727 (or to require only a small part of the spectrum). However they depend on fainter lines than $R_{23}$ and tend to have larger scatter on their calibrations.

Another characteristic of the strong line methods is that they have to be calibrated by a different method. As such, any determination made by the strong line methods will have the uncertainties associated with the scattering within the particular strong line method, as well as with the biases and scattering associated with the method used for calibration. Specifically it does not provide a solution to the ADF problem, but rather the ADF has to be properly understood, so that any strong line method can be properly calibrated, before accurate abundance determinations can be made.

As it stands today different calibrations of Pagel's method vary by nearly an order of magnitude \citep{Kewley2008}; however, most of this scatter disappears when one uses a calibration that also considers the degree of ionization, to make a more robust determination. Yet, even after that determinations calibrated using CELs and the direct method \citep[e.g.][]{Pilyugin2005} differ from those using RLs or CELs + $t^2$ \citep[e.g.][]{PenaGuerrero2012a} by a factor of about 2.

\subsection{Element depletions}\label{ssec:total_depletions}

The abundances derived from emission lines following the steps described before provide information about the chemical composition of the ionized gas, but elements can be also deposited in dust grains. The underabundance found for many elements with respect to a reference value is generally interpreted as due to depletion onto dust grains. The depletion factor of an element $X$ is given by:
\begin{equation}
[X/{\rm H}] = \log(X/{\rm H})_{\rm gas} - \log(X/{\rm H})_{\rm true},
\end{equation}
where $\log(X/{\rm H})_{\rm gas}$ is the observed gaseous abundance and $\log(X/{\rm H})_{\rm true}$ is the true abundance of the object; for many practical purposes the true chemical abundances are not known, instead a ``cosmic'' (or reference) abundance can be used. The composition of the Sun is often taken as the reference composition but other abundances can be used, like those of young O-  and B-type stars.

Noble gases such as helium, neon, and argon are not expected to be important contributors to the dust, volatile elements like carbon and oxygen are somewhat embedded in grains, and refractory elements such as iron and calcium are mostly deposited in the dust \citep{Whittet2003}. The elements that contribute most to the mass of dust grains are carbon, oxygen, magnesium, silicon, and iron \citep[e.g.,][]{Whittet2010}. A review on element depletion for 17 elements has been done by \citet{Jenkins2009}.

Generally, element depletions are only used to correct the gaseous abundances and compute total element abundances. For example, to compare the chemical compositions of PNe and \Hiiregs\ with those obtained in other sets of objects. However, depletions can also be used to gain information about dust formation and evolution in different environments. Some examples of these two types of studies are presented in the next two sections. 

\section{Recent results from \ion{H}{2} regions} \label{sec:important_HII}

\subsection{Abundance discrepancy factor in \Hiiregs} \label{ssec:ADF_HII}

There are three different methods to determine abundances in \Hiiregs\ that are frequently used: a) the strong lines method, that is based on the intensity ratio of the nebular lines (such as those of [\ion{O}{2}], [\ion{O}{3}], [\ion{N}{2}], [\ion{S}{3}], and [\ion{Ne}{3}]), relative to those of hydrogen, this method is calibrated with a net of photoionized models such as those provided by CLOUDY \citep{Ferland2013}, b) the direct method, that is based on the ratio of forbidden lines to hydrogen lines, and the temperature derived from the ratio of nebular lines to auroral lines like those of [\ion{O}{3}], [\ion{N}{2}], [\ion{S}{3}], and c) the RLs method that is based on the intensities of the RLs of C, O, and helium  relative to those of hydrogen, that is almost independent of the \temp. The intensity of the nebular lines of [\ion{O}{3}] is typically two orders of magnitude higher than that of the auroral lines, and three orders of magnitude higher than that of the \ion{O}{2} RLs. Therefore the telescope time to derive the intensity of the required lines increases considerably when  going from method a) to method c).

There is a long-standing riddle when applying $T_{\rm e}$ for the determination of abundances relative to hydrogen. One generally derives higher relative abundances from RLs than from the forbidden lines. This is called the abundance discrepancy factor (ADF) problem, and a possible origin lies in temperature inhomogeneities, parameterized as $t^2$ \citep{Peimbert1967, Peimbert1993}. The presence of temperature inhomogeneities causes the abundances determined from the forbidden lines to be underestimated, accounting for the ADF.

The ADF in gaseous nebulae can be due to inhomogeneities in temperature, density, and chemical composition. Gaseous nebulae indeed show density and temperature inhomogeneities and some of them also show chemical inhomogeneities. Therefore these three possible causes for the ADF values should be studied to determine the real abundances in gaseous nebulae. Temperature inhomogeneities are also known as temperature fluctuations or temperature variations in the literature.

\subsection{The $\kappa$ electron distribution} \label{ssec:kappa}

It has been recently proposed that the free electrons in \Hiiregs\ and PNe might have significant deviations from a Maxwellian velocity distribution, due to the presence of supra thermal electrons, and that their distribution can be represented by a generalized Lorentzian distribution given by the $\kappa$ formalism \citep{Nicholls2012, Nicholls2013, Dopita2013}. \citet{Ferland2016} have shown that the distance over which heating rates change are much longer than the distance supra-thermal electrons can travel, and that the timescale to thermalize these electrons is much shorter than the heating or cooling timescales. These estimates imply that supra-thermal electrons will have disappeared into the Maxwellian velocity distribution long before they affect the collisionally-excited forbidden and RLs, therefore the electron velocity distribution will be closely thermal and the $\kappa$ formalism can be ruled out for these objects.

\subsection{Temperature inhomogeneities in \Hiiregs} \label{ssec:Tvar_HII}

The difference between the O/H values obtained from oxygen forbidden lines and those derived from RLs can be due to the presence of temperature inhomogeneities in \Hiiregs. The importance of the temperature inhomogeneities can be estimated by means of $t^2$; which can be estimated by comparing temperatures derived from two different methods, for example: a) by comparing the $T$ derived form the ratio of the Balmer continuum to a Balmer line with the temperature derived from the ratio of two forbidden lines, b) by comparing the temperature derived from \ion{He}{1} RLs with a temperature derived using forbidden lines, also c) by comparing the abundances, for the same ion, derived using RLs with those derived using CELs. From a group of 37 galactic and extragalactic \Hiiregs\ observed by different authors \citet{Peimbert2012} found $t^2$ values in the 0.019 to 0.120 range with an average value of 0.044 based on the \ion{O}{2} RLs and the [\ion{O}{3}] CELs. The $t^2$ average value is considerably higher than the Orion nebula value that amounts to $0.028 \pm 0.006$.

\subsection{Chemical Inhomogeneities in \Hiiregs} \label{ssec:ChemicalInhomogeneities_HII}

In a chemically inhomogeneous medium, CELs are expected to originate mainly in regions that are relatively metal-poor, temperature-high, and density-low, while the RLs are expected to originate mainly in regions that are relatively metal-rich, temperature-low, and density-high. Based on high-quality observations of multiplet V1 of \ion{O}{2} and the NLTE atomic computations of  \ion{O}{2}, \citet{Peimbert2013} study the density and temperature of a sample of  \ion{H}{2}  regions. They  find that the signature for oxygen-rich clumps of high density and low temperature is absent in all objects of their sample: one extragalactic and eight Galactic  \ion{H}{2}  regions. The temperatures derived from: (1) RLs of  \ion{O}{2}, and (2) RLs of  \ion{H}{1}  together with Balmer continua are lower than those derived from forbidden lines, while the densities derived from RLs of  \ion{O}{2}  are similar or smaller than densities derived from forbidden lines. Electron pressures derived from CELs are about two times larger than those derived from RLs. These results imply that the proper abundances are those derived from RLs and that these nebulae are chemically homogeneous

\subsection{The Orion Nebula } \label{ssec:O/H_gas}

\subsubsection{The gaseous O/H value in the Orion Nebula } \label{sssec:O/H_gas}

The gaseous O/H value in \Hiiregs\ can be obtained from CELs or RLs of oxygen relative to RLs of H. 

The Orion Nebula is the best observed \ion{H}{2} region. Orion is in the plane of the Galaxy, about 400 pc farther away from the nucleus of the Galaxy than the Sun. There have been many O/H gaseous abundance determinations of the Orion nebula based on oxygen forbidden lines. Representative values derived during the last 40 years by different authors, in 12 + log(O/H) units, are: 8.52 \citep{Peimbert1977}; 8.49 \citep{Osterbrock1992}; 8.51 \citep{Peimbert1993}; 8.47 \citep{Esteban1998}; 8.51 \citep{Deharveng2000}; 8.51 \citep{Esteban1998, Esteban2004}; results that have been nearly constant through four decades and several atomic physics calculations. From these results we estimate that: 12 + log(O/H) = $8.50\pm0.02$. Alternatively based on RLs \citet{Esteban1998,Esteban2004} obtain that: 12 + log(O/H) = $8.72\pm0.07$ and $8.71\pm0.03$ respectively. 

The difference between both types of determinations is real and an explanation for this difference has to be sought. The abundances derived from the ratio of two RLs are almost independent of the temperature structure, while the abundances derived from the ratio of a forbidden line to a recombination line depend strongly on the temperature structure. 

In the presence of temperature inhomogeneities over the observed volume, and adopting the temperatures derived from the ratio of two forbidden lines, the O/H abundances derived from oxygen forbidden lines become smaller than the real ones. While the O/H abundances derived from RLs are almost independent of the temperature structure and therefore are representative of the real O/H values.

The difference between the abundances derived from forbidden lines and those derived from RLs has been called the abundance discrepancy factor, ADF. See section \ref{sssec:ADF_PNe} for further discussion on the ADF.

\subsubsection{Temperature inhomogeneities in the Orion Nebula} \label{sssec:Tvar_Orion}

The difference between the O/H values obtained from oxygen forbidden lines and those derived from RLs can be due to the presence of temperature inhomogeneities in \Hiiregs. The importance of the temperature inhomogeneities can be estimated by the mean squared temperature fluctuation, $t^2$, frequently named the $t^2$ value or $t^2$ parameter, or $t^2$. $t^2$ can be estimated by comparing temperatures derived from two different methods, for example: a) by comparing the \temp\ derived from the ratio of the Balmer continuum to a Balmer line with the temperature derived from the ratio of two forbidden lines, b) by comparing the temperaature derived from the ratios of a large number of \ion{He}{1} RLs, that have slightly different temperature dependence, with the temperature derived from the ratio of two forbidden lines, c) by comparing the intensities of a recombination, a nubular, and an auroral line originating from the same ion.

\subsubsection{The dust contribution to the total O/H ratio in the Orion Nebula} \label{sssec:dust_O/H}

In this branch of astrophysics it is relatively easy to determine the ratio of oxygen to hydrogen atoms in the photoionized gas, an obvious followup of this study is to try to compare these determinations with those made in other sets of objects, specially since oxygen amounts to $\sim 50\%$ of the heavy elements. Unfortunately, the gaseous phase in \Hiiregs\ is not a good representation of the chemical composition of the object.

Recently, \citet{Espiritu2017} found that the fraction of oxygen atoms embedded in dust in the Orion nebula amounts to $0.126\pm0.024$ dex in agreement with the previous value found by \citet{MesaDelgado2009}.

Accurate abundance corrections due to depletion have only been done for oxygen. While depletion is not important for noble gases (He, Ne, and Ar), depletion of metallic elements (Fe, Ni, Mg, etc.) is so high that studies of the gas phase will not permit us to determine the true abundances of such atoms. Finally, depletions of elements such as C, N, S, or Cl have not been studied in any detail and a depletion of 10 - 40\% is to be expected, so comparison between stellar and nebular abundances of such elements should be taken with a grain of salt.

\subsubsection{Comparison of the total O/H value in the Orion nebula with the O/H value of the B stars in the Ori OB1 association} \label{sssec:O/H_Orion}

\citet{Nieva2011} have determined the O/H ratio for 13 B stars of the Ori OB1 association and obtain a 12 + log(O/H) = $8.77\pm0.03$ value. This value has to be compared with the log O/H value (gas+dust) in the Orion nebula that amounts to $8.84 \pm 0.04$. These results are in fair agreement and imply that the total O/H value for the Orion \ion{H}{2} region and their associated B stars is about 12 + log(O/H) = $8.80 \pm 0.05$. A value considerably higher than the O/H gas ratio in the Orion nebula derived from forbidden lines and the assumption of $t^2$ = 0.00 that amounts to 12 + log(O/H) = 8.50.

\subsubsection{Comparison of the Orion nebula O/H values with those of the solar vicinity} \label{ssec:O/H_Orion.Sun}

\citet{Asplund2009} find a protosolar 12 + log(O/H) = $8.73\pm0.05$ value. To compare the solar value with the Orion nebula value we have to consider the chemical evolution of the Galaxy during the last 4.6 Gyr, and the galactocentric distance at which the Sun was formed. \citet{Carigi2011} from a chemical evolution model of the solar vicinity, have estimated that the Sun originated at a similar galactocentric distance than the one it has now.

The present O/H ratio in the solar vicinity is 0.02 dex higher than in the Orion association, because the O/H gradient of the Insterstellar Medium (ISM) amounts to 0.044 dex per kpc \citep{Esteban2005}. In the solar neighborhood, according to models of galactic chemical evolution, the O/H ratio during the last 4.6 Gigayears has increased by 0.13 dex \citep{Carigi2011}. By adding these two values to the protosolar value we expect for a recently formed star in the solar vicinity a 12 + log(O/H) = 8.88 value. From the Orion nebula we expect a 12 + log(O/H) = 8.84 value for the solar vecinity, in fair agreement with the previous estimate.

There are two other estimates of the O/H value in the ISM that can be made from observations of F and G stars of the solar vicinity. According to \cite{Allende-Prieto2004} the Sun appears deficient by roughly 0.1 dex in O, Si, Ca, Sc, Ti, Y, Ce, Nd, and Eu, compared with its immediate neighbors with similar iron abundances; by assuming that the oxygen abundances of the solar immediate neighbors are more representative of the present day local ISM than the solar one, and by adding this 0.10 dex difference to the oxygen photospheric solar value by \citet{Asplund2009}, that amounts to 8.69, we obtain that the present value of the ISM has to be higher than 12 + log(O/H) = 8.79. A similar result is obtained from the data by \cite{Bensby2006} who obtain for the six most O-rich thin-disk F and G dwarfs of the solar vicinity an average O/H value 0.16 dex higher than the solar photospheric value; by assuming their value as representative of the present day ISM of the solar vicinity we find 12 + log(O/H) = 8.85. Both results are in good agreement with the O/H Orion nebula value derived from the \ion{O}{2} RLs method, after adding the fraction of oxygen embedded in dust grains.

\subsection{The primordial helium abundance, $Y_P$} \label{ssec:YP}

According to big bang nucleosynthesis (BBN) the primordial abundances of helium and hydrogen by unit mass, $Y_P$ and $X_P$, produced during the first four minutes after the start of the expansion of the universe amount to  about 0.25 and 0.75 respectively  \citep[e. g.][and references therein]{Peimbert2008,Pagel2009,Cyburt2016}. The best objects to determine the primordial helium  and hydrogen   abundances are galaxies where star formation has been very scanty and therefore present almost no heavy elements. 

A thorough discussion on the open problems related to the primordial light element abundances was presented during the IAU Symposium: Light Elements in the Universe \citep{Charbonnel2010}. In particular, the subject of the primordial helium abundance was reviewed on a round table discussion \citep{Ferland2010}.

The three most recent determinations of $Y_P$ based on observations of \Hiiregs\ in metal poor galaxies are those by \citet{Izotov2014}, \citet{Aver2015}, and \citet{Peimbert2016}. Their results are presented in Tables  \ref{tab:neutrino} and \ref{tab:neutron}.

The determination of $Y_P$ based on BBN depends on several input values, like the number of neutrino families $N_{\nu}$ and and the neutron  lifetime $\tau_n$. From the observed $Y_P$ value, assuming BBN and adopting $\tau_n=880.3$s \citep{Olive2014} it is possible to determine $N_{\nu}$, see Table  \ref{tab:neutrino}. Alternatively by adopting an $N_{\nu}$ of 3.046 \citep{Mangano2005} it is possible to determine $\tau_n$, see Table \ref{tab:neutron}. The $Y_P$ values by \citet{Aver2015} and \citet{Peimbert2016} agree with each other but  are in disagreement with the value by \citet{Izotov2014} by more than $3\sigma$.

\begin{deluxetable}{lcc}
\tablecaption{$Y_P$ values and predicted equivalent number of neutrino families, $N_{\nu}$, assuming $\tau_{n}=880.3$ s.
\label{tab:neutrino}}
\tablehead{
\colhead{$Y_P$} & 
\colhead{$N_{\nu}$\tablenotemark{a}} &
\colhead{$Y_P$ source}
}
\startdata
 $0.2551\pm0.0022$ & $3.58\pm0.16$ & Izotov et~al.\@ (2014) \\
 $0.2449\pm0.0040$ & $2.91\pm0.30$ & Aver et~al.\@  (2015) \\
 $0.2446\pm0.0029$ & $2.89\pm0.22$ & Peimbert et al.\@ (2016)  \\
\enddata
\end{deluxetable}

\begin{deluxetable}{lcc}
\tablecaption{$Y_P$ values and the neutron mean life, \lowercase{$\tau_{ n}$}\label{tab:neutron}, assuming $N_{\nu}=3.046$.}
\tablehead{
\colhead{$Y_P$} & 
\colhead{$\tau_{n}$(s)\tablenotemark{a}} &
\colhead{$Y_P$ source}
}
\startdata
 $0.2551\pm0.0022$ & $921\pm11$ & Izotov et~al.\@ (2014) \\
 $0.2449\pm0.0040$ & $872\pm19$ & Aver et~al.\@  (2015) \\
 $0.2446\pm0.0029$ & $870\pm14$ & Peimbert et al.\@ (2016)  \\
\enddata
\end{deluxetable}

\subsection{Dust in \Hiiregs} \label{ssec:Dust_HII}

To compare the chemical abundances of \Hiiregs\ with stellar abundances it is necessary to add to the gaseous abundances in \Hiiregs\ the fraction of the elements embedded in dust grains. The fractions of H, He, Ne and Ar in dust grains  are expected to be negligible; on the other hand, the fractions of Mg, Si, and Fe, as well as for most true metals, are expected to be so large that the gas phase abundance is not usefull to determine the true chemical abundance; between these 2 extremes, elements like C, N, O, S and Cl are expected to be mostly in gasous form, but with a substantial fraction of their atoms trapped in dust grains.

From the study of Galactic and extragalactic \Hiiregs, \citet{Rodriguez2005} found that the iron depletion seems to increase at higher metallicities. In Table 3 we present the O, Mg, Si, and Fe depletions derived for the Orion nebula and 30 Doradus by \citet{Peimbert2010}. From the study of 78 \Hiiregs\ they find that the fraction of Fe embedded in dust grains increases with the total O/H ratio. 

\begin{deluxetable}{lcc}
\tablecaption{Fraction  (\%) of atoms enbedded in dust grains in \Hiiregs \label{tab:dust}}
\tablehead{
\colhead{Element} & 
\colhead{Orion} &
\colhead{30 Doradus}
}
\startdata
O   & $24\pm5$         & $22\pm5$           \\
Mg & $91\pm3$         & $72\pm9$           \\
Si  & $76\pm7$          & $81^{+16}_{-9}$ \\
Fe & $97^{+1}_{-2}$  & $92^{+4}_{-3}$   \\
\enddata
\end{deluxetable}

\citet{Peimbert2010} based on the depletions of Mg, Si, and Fe atoms, estimated that in \Hiiregs\ the fraction of oxygen atoms embedded in dust grains increases from about 0.08 dex, for the metal poorest \Hiiregs\ known, to about 0.12 dex, for metal rich \Hiiregs. Recently \citet{Espiritu2017} found that in the Orion nebula $25\pm4\%$ of the oxygen atoms are embedded in dust.

\subsection{Calibration of Pagel's method to determine the O/H abundances} \label{ssec:Pagel}

\citet{Pagel1979} found that the $I$([\ion{O}{2}] 3727 + $I$[\ion{O}{3}] 4959 + 5007)/$I$(H$\beta$) versus $n$(O)/$n$(H) diagram varied smoothly with the total O/H abundance ratio and proposed to use it to determine the O/H ratio for objects where the intensity of the [\ion{O}{3}] $\lambda$4363 line, needed to derive the temperature, is not available. This diagram now is known as the $R_{23}$ - O/H diagram. \citet{PenaGuerrero2012a} calibrated the $R_{23}$ - O/H diagram taking into account the fraction of oxygen trapped in dust grains and the presence of temperature inhomogeneities in \Hiiregs. Based on 28 \Hiiregs\ they find that these two corrections increase the  O/H vaues by 0.22 - 0.35 dex relative to the gaseous O/H values assuming constant temperature.

\subsection{More on abundances of Galactic and extragalactic \Hiiregs} \label{ssec:EG_Gradients}

One of the problems of deriving abundances relative to hydrogen is the dependence on the temperature of the forbidden lines in the visual region. There are three ways to try to diminish the importance of the \temp\ on the abundance determinations: a) to derive the abundances from RLs relative to hydrogen, b) to use CELs in the infrared that depend weakly on the temperature, c) to obtain the ratio of two forbidden lines of different elements from collisionally excited levels of similar energy.

\subsubsection{Our Galaxy} \label{sssec:MW_HII}

\citet{Esteban2005} present C/H and O/H abundances for eight \Hiiregs\ in our galaxy based on RLs and obtain well defined galactocentric abundance gradients. \citet{Carigi2005} produce a chemical evolution model of the galaxy that fits  the observed carbon, nitrogen, and oxygen galactic gradients and find that about half of the carbon in the ISM has been produced by massive stars and half by low and intermediate mass stars. \citet{Carigi2011} present chemical evolution models that fit a) the observed C/H and O/H galactocentric gradients obtained from \Hiiregs\ b) the hydrogen, helium, carbon, and oxygen abundances of M17, c) the protosolar abundances, and d) the C/O, O/H, and C/Fe relations derived from solar vicinity stars.

\citet{Esteban2013}, based on RLs have determined the abundances of carbon and oxygen  for  the Galactic \ion{H}{2} region NGC 2579; they find that the chemical composition of this \ion{H}{2} region is consistent with flattened carbon and oxygen  gradients at its galactocentric distance of $12.4\pm0.7$ kpc, they also find  that a levelling out of the star formation efficiency about and beyond the isophotal radius of the Galaxy can explain the flattening  of the chemical gradients, they also obtain that  $t^2 = 0.045\pm0.007$ for NGC~2579, a value similar to those obtained for other galactic and extragalactic \Hiiregs.

\subsubsection{Other spiral galaxies} \label{sssec:OtherSpiral_HII}

\citet{Esteban2009} derive C/H and O/H abundance gradients based on RLs for M33, M101, and NGC 2403; the C/H gradient is steeper than the O/H gradient. This result is similar to that found in the Milky Way and has important implications for chemical evolution models of spiral galaxies. \citet{Zurita2012} based on the direct method (direct $T_e$-based method) find that the Ne/O, Ar/O, and S/O abundance ratios are consistent with a constant value across the M31 disk. \citet{Croxall2013} present observations of the [\ion{O}{3}] 88 micrometers fine structure line, which is insensitive to temperature, they combine these data  with optical observations to obtain oxygen abundances, these abundances are in agreement with estimates that assume that temperature inhomogeneities are present in \Hiiregs. 

\citet{Berg2015} based on the direct method obtain abundances for 45 \Hiiregs\ in the spiral galaxy NGC 628, the S/O, Ne/O and Ar/O are constant across the galaxy. \citet{ToribioSanCipriano2016} from RLs of \ion{O}{2} and \ion{C}{2} in \Hiiregs\  derive C/H and O/H gradients for NGC 300 and M33, the C/H gradients are steeper than those of O/H leading to negative C/O galactocentric gradients. \citet{Bresolin2016} find that oxygen RLs yield nebular abundances that agree with stellar abundances for high-metallicity systems, but find evidence that in more metal poor environments the oxygen RLs tend to be higher than those derived from stellar abundaces. \citet{Croxall2016} present Large Binocular Telescope observations for 109 \Hiiregs\ in M101, for 74 of them  they determine abundances by the direct method; they study different ICFs in the literature and find a group of \Hiiregs\ with low Ne/O ratios that defies explanation.

\citet{ToribioSanCipriano2017} have studied five \Hiiregs\ in the Large Magellanic Cloud and four in the Small Magellanic Cloud. They derive the O/H, C/H, and C/O based on RLs. They find that the LMC seems to show a similar chemical evolution to that of the external zones of small spiral galaxies and that the SMC behaves as a typical star-forming dwarf galaxy. By comparing the nebulae and B-type stellar abundances they find that they agree better with the nebular ones derived from CELs. Comparing the results with other galaxies they find that the stellar abundances seem to agree better with the nebular abundances derived from CELs, in low metallicity environments and from RLs in high metallicity environments; this behaviou  suggests that either a) there are 2 processes creating the ADFs, one afecting RLs (mostly at low metallicities) and one afecting CELs (mostly at high metalicities), or b) there are problems in the abundance determinations of B-type stars (either at low- or high-metalicities).

\subsubsection{Irregular galaxies} \label{sssec:Irregular_HII}

\citet{Esteban2014} have studied \Hiiregs\ in star forming dwarf irregular galaxies, they find that the \Hiiregs\  occupy a different locus in the C/O versus O/H diagram than those belonging to the inner disks of spiral galaxies, indicating their different chemical evolution histories, and that the bulk of carbon in the most metal poor extragalactic \Hiiregs\ should have the same origin than in halo stars. The comparison between the C/O ratios in these \Hiiregs\ and stars of the Galactic thick and thin disks seems to give arguments to support the merging scenario for the origin of the Galactic thick disk.

\citet{Berg2016}, based on Hubble Space Telescope observations of 12 \ion{H}{2} regions in low metallicity dwarf galaxies, determined the C/O abundance ratio from the UV \ion{C}{3} $\lambda\lambda \, 1906+1909$ and the \ion{O}{3}] $\lambda1666$ lines ratio, this ratio is almost independent of dust absorption and the adopted \temp. They find that there is no increase of C/O with O/H at low metallicities, at higher O/H values there seems to be a general increasing trend of C/O versus O/H. They also find that the C/N ratio appears to be constant over a large range in the O/H abundance.

\section{Recent results from Planetary Nebulae} \label{ssec:important_PNe}

\subsection{Abundance discrepancy factor in PNe} \label{sssec:ADF_PNe}

The ADF in PNe can be due to inhomogeneities in temperature, density, and chemical composition. Gaseous nebulae indeed show density and temperature inhomogeneities (also known as fluctuations or variations) and some of them also show chemical inhomogeneities. Therefore these three possible causes for the ADF values should be studied to determine the real abundances in a given PN. 

\subsection{Density inhomogeneities in PNe} \label{sssec:Dens-Var_PNe}

The effect of electron densities on the abundance determinations of gaseous nebulae can also mimic spurious ADF values. \citet{Rubin1989}, \citet{Viegas1994}, and \citet{Tsamis2011} have studied the dependence of the line intensities on density when the upper energy levels producing forbidden lines are de-excited by collisions. The critical density for collisional de- excitation is different depending on the line of a given ion. If this effect is not taken into account, the temperatures derived from CELs are overestimated and the abundances are underestimated. This can be the case for high-density gaseous nebulae and for certain ions. This effect is particularly relevant when infrared lines are used to determine abundances and for objects of relatively high density. We do not expect this effect to be important for the determination of the O$^{++}$/H$^+$ ratio, since  the $\lambda4363$ and $\lambda4959$ [\ion{O}{3}] lines have critical densities of $2.4 \times 10^7 {\rm cm}^{-3}$ and $6.4 \times 10^5 {\rm cm}^{-3}$, respectively, values that are considerably higher than the densities of the typical \Hiiregs\ and PNe. 

\subsection{Temperature and Chemical Inhomogeneities in PNe} \label{sssec:Temp-Chem-Var_PNe}

\citet{Georgiev2008} found that the central star of NGC 6543 is not H-poor, and has a normal helium  composition. Moreover they also found that the $t^2$ determined by five different methods indicates that the difference in nebular abundances between the RLs and CELs can be explained as due to temperature inhomogeneities in a chemically homogeneous medium. These  results  are in disagreement with those by \citet{Wesson2004}, who propose that NGC 6543 contains high-density H-poor inclusions that are rich in helium and heavy elements.

\citet{Liu2006} presented a review giving arguments in favor that the ADF in PNe is mostly due to chemical inhomogeneities, while \citet{Peimbert2006} presented a review giving arguments that the ADF in most planetary nebulae is due to temperature inhomogeneities in chemically homogeneous objects.

It has been suggested that the ADF problem can be due to the destruction of solid bodies inside PNe that produce cool and high-metallicity pockets \citep{Henney2010}.

\citet{Liu2000} concluded that temperature fluctuations in NGC~6153 cannot account for its ADF. \citet{McNabb2016} have studied three objects with very large ADF values and conclude that NGC 6153 is chemically inhomogeneous. While \citet{Peimbert2014} present results for 16 PNe and conclude that NGC 6153 is chemically homogeneous. The difference between such conclusions arises because McNabb et al. find very low recombination temperatures ($\sim3000$ K) while Peimbert et al. find moderately low recombination temperatures ($\sim6000$ K) and moderately low recombination densities ($\sim$5500 cm$^{-3}$), incompatible with the presence of high-density/low-temperature blobs.

Recently, \citet{Storey2017} have computed the recombination coefficients for O$^{++}$ + e$^{-}$ using an intermediate coupling treatment that fully accounts for the dependence of the distribution of population among the ground levels of O$^{++}$ on \dens\ and \temp. Based on these atomic data it is possible to obtain, from high quality observations of the \ion{O}{2} recombination lines, the N(O$^{++}$)/N(H$^{+}$), \dens\ and \temp\ values. By combining these results with the values derived from forbidden lines of [\ion{O}{2}] it is possible to study the presence of temperature, density, and abundance inhomogeneities. Storey et al. present results for densities and temperatures based on the \ion{O}{2} lines available in the literature and show that  high ADF PNe present considerably lower temperatures than low ADF PNe and \ion{H}{2} regions.

Further discussion on temperature and chemical inhomogeneities is presented in section 7.

\subsection{Ionization correction factors for PNe} \label{sssec:ICF_PNe}

In the last years, new ICFs based on grids of {\sc cloudy} photoionization models have been computed for He, C, N, O, Ne, S, Cl, and Ar \citep{DelgadoInglada2014a}, Zn \citep{Smith2014}, Se and Kr \citep{Sterling2015}, and Ni \citep{DelgadoInglada2016b}. 

\subsection{Element production by PN progenitors} \label{sssec:Prod_PNe}

The progenitor stars of PNe may go through several nucleosynthesis processes during their lives that alter their initial chemical composition. The mechanisms that take place and their efficiencies depend on the initial mass of the star and the metallicity, for example, they are generally more efficient at lower metallicities. In brief, all the AGB stars produce some helium ($^4$He) and nitrogen ($^{14}$N), and destroy some carbon ($^{12}$C) via the first dredge-up (FDU). The second dredge-up (SDU) takes place in stars with masses above 3.5--4 $M_\odot$ (depending on the metallicity and the model) and its major effect is a significant increase in the helium abundance. The third dredge-up (TDU) may occur in stars with masses above 1--1.5 $M_\odot$ and it is responsible for the carbon stars. Thermal pulses and dredge-ups bring to the surface carbon and s-process elements. Finally, the hot bottom burning (HBB) occurs in the most massive PN progenitors, those with masses above 3.5--4 $M_\odot$ (those stars where the SDU occurs). The net result of the HBB is a significant increase of nitrogen and the destruction of carbon (in the most massive AGB stars oxygen may also be destroyed). A thorough discussion on nucleosynthesis processes can be found in the review by \citet{Karakas2014}. 

The comparison between PN abundances computed from observations and the theoretical predictions from stellar nucleosynthesis models is necessary to: a) understand the nucleosynthesis that occur in low- and intermediate-mass stars, and b) their impact on the chemical enrichments of galaxies \citep[see, e.g.,][]{GarciaHernandez2014, DelgadoInglada2015, GarciaHernandez2016a, GarciaHernandez2016b, GarciaRojas2016b}. 

Traditionally, the PNe with high abundances of helium and nitrogen are associated with high mass progenitors; these PNe are called Type I PNe according to the classification made by \citet{Peimbert1978}. However, recent results have shown that extra mixing processes such as stellar rotation, thermohaline mixing, magnetic fields and other mechanisms could explain helium  and nitrogen  large abundances in low mass stars that probably have not gone through the SDU and the HBB \citep{Stasinska2013, Karakas2016}.    

\cite{DelgadoInglada2015} found evidences of oxygen production in some PN progenitor stars. An oxygen enrichment of $\sim0.3$ dex is present in some Galactic PNe that arise from stars with masses $\sim$1.5 M$_{\odot}$ formed at sub-solar metallicity. This result can be explained as a consequence of oxygen being produced through the nuclear burning of helium and then dredge-up to the stellar surface via the TDU. The efficiency of this mechanism depends on the mass of the star and the metallicity, and it varies from model to model. Only the stellar models that include an extra-mixing process, such as those computed by \citet{Ventura2013} and \citet{Pignatari2016}, are able to reproduce the observations. 

The abundances of several $n$-capture element abundances (such as  P, F, Ge, Se, Br, Kr, Rb, Ca, and Xe) have been recently computed using high-resolution and deep spectra with new ICFs \citep{Otsuka2011, Otsuka2013, GarciaRojas2015, Sterling2015, Sterling2016}. Also a clear correlation has been found between the C/O values in PNe and the $n$-capture element enrichment; this indicates that $n$-capture elements and carbon are brought together to the surface in the same mechanism, the TDU \citep[see, e.g.,][]{Sterling2015}. Moreover, it has been found that stars with different masses are enriched in a different amount in $n$-capture elements \citep[see, e.g.,][]{Sterling2015}.

Some interesting recent reviews on chemical abundances of PNe were made by \citet{Kwitter2012, Magrini2012, DelgadoInglada2016a, GarciaRojas2016}.

\subsection{Dust in PNe} \label{sssec:Dust_PNe}

In our Galaxy, AGB stars dominate dust production over SNe and, in general, these stars are one of the most efficient sources of dust in galaxies \citep{Whittet2003}. As the leftovers of many AGB stars, PNe are adequate places to study the life cycle of dust grains in the circumstellar gas. The presence of dust in PNe is evidenced by the strong infrared excess produced by the emission from dust grains (\citealt{Balick1978, Kwok1980}). In addition, the gaseous abundances of refractory elements (such as Mg, Al, Si, Ca, Fe, and Ni) in PNe are much lower than the solar values and this is generally attributed as due to their depletion onto dust grains \citep[see, e.g.,][and references therein]{DelgadoInglada2009}. 

The depletion factors of iron and nickel in Galactic PNe have been recently studied by \citet{DelgadoInglada2014b} and \citet{DelgadoInglada2016b} using optical spectra and ICFs. They obtained that, in most of the PNe, up to $\sim$90\% of the Ni and Fe atoms are condensed into dust grains. \citet{DelgadoInglada2016b} also found that: 1) iron atoms tend to be more deposited in grains in PNe with C-rich dust than in PNe containing O-rich dust and 2) nickel atoms tend to be more attached to the grains than iron atoms in environments with a higher depletion. 

As we mentioned in Section~5.7, that the iron depletion in \Hiiregs\ increases as the metallicity increases \citep{Rodriguez2005, Peimbert2010}. To perform a similar analysis with PNe, deep spectra with high resolution of PNe at low metallicities are needed.

Another approach to study the dust present in PNe is through their infrared dust features. Traditionally, PNe are believed to have either carbon-rich dust or oxygen-rich dust. The grains are formed in the cool and dense atmospheres of AGB stars. One expects that stars with C/O $>1$ in their envelopes would form carbon-rich grains whereas those with C/O $<1$ would form oxygen-rich grains. The value of C/O in the surface of the star depends on the nucleosynthesis that has occurred during the AGB phase. PNe arising from oxygen-rich AGB stars are expected to have oxygen-rich dust (such as oxides and silicates) and those coming from carbon-rich AGB stars would contain molecules associated with carbon-rich chemistry (such as SiC, MgS, and polycyclic aromatic hydrocarbons, PAHs). 

Since the finding of the first carbon-rich star containing oxygen-rich dust \citep{Waters1998}, several PNe have been found that show at the same time PAHs (C-rich molecules) and amorphous and/or crystalline silicates (\citealt{PereaCalderon2009, DelgadoInglada2014b, GarciaHernandez2014, GuzmanRamirez2014, Cox2016}; Garc\'ia-Rojas et al. 2017, in prep.). These objects are called PNe with double chemistry dust and the reason for this duality is still a mystery. Some authors suggest that AGB stars may suffer a late thermal pulse that changes the stellar outflow from O-rich into C-rich, making possible for these stars to produce both oxygen- and carbon-rich dust \citep[see, e.g.,][]{PereaCalderon2009}. In this scenario the observed C/O in the PNe is expected to be greater than 1. On the other hand, it has been proposed that the CO molecules may be dissociated by ultraviolet photons in a oxygen-rich environment and the free carbon atoms could be used to form carbon-rich dust \citep[see, e.g.,][]{GuzmanRamirez2014}. This could explain the observation of PAHs in O-rich PNe (C/O $< 1$). Further analysis is needed to solve this question.

\subsection{Galactic and extragalactic O/H gradients from PNe and \Hiiregs} \label{sssec:Extragalactic_PNe}

\citet{Stanghellini2010} have determined the O/H gradient for our galaxy using the direct method for PNe of types I, II, and III defined by \citet{Peimbert1978}. The ages and initial masses of the PNe progenitors are: Type I ages smaller than about 1 Gyr and initial masses $\gtrsim$2 $M_\odot$, type II ages in the $\sim 1-5$ Gyr range and masses in the $\sim 1.2-2.0 M_\odot$ range, and Type III ages $\gtrsim$5 Gyr and initial masses $\lesssim$1.2 $M_\odot$. In Table~\ref{tab:grad} we present the O/H gradients derived by \citet{Stanghellini2010}, and we also show the O/H gradient derived for \Hiiregs\ based on recombination lines by \citet{Esteban2005}. \Hiiregs\ have ages of at most a few million years. Table~\ref{tab:grad} shows that there is a clear trend in the sense that the older the set of objects the flatter the derived gradient.

In a study of photoionized regions in M81, \citet{Stanghellini2014}, based on the direct method, have obtained for \Hiiregs\ an O/H gradient of $-0.088\pm0.013$ dex kpc$^{-1}$ and for PNe an O/H gradient of $-0.044\pm0.007$ dex kpc$^{-1}$. These results can be interpreted as due to a flattening of the interstellar O/H gradient with time. \citet{Magrini2016}, based on the direct method have derived the PNe and \Hiiregs\ gradients for NGC 300, M 33, M31, and M81. For these spiral galaxies the PN gradients are flatter or equal to those of \Hiiregs. The O/H interstellar values increase with time, moreover M31 and M81 show larger increments with time in the (\ion{H}{2}-PN) O/H ratios at a given galactocentric distance than M33 and NGC 300. \citet{Magrini2017} find that NGC 55 shows no trace of radial gradients, they discuss the differences between this galaxy and NGC 300 that does show an O/H gradient.

\begin{deluxetable}{lcc}
\tablecaption{Radial metallicity gradients in the Milky Way given in dex/kpc. \label{tab:grad}}
\tablehead{\colhead{Objects} & \colhead{$\Delta$log(O/H)/$\Delta$R} & \colhead{Source}}
\startdata
PNe all & $-0.023\pm0.006$ & (1) \\
PNe Type I & $-0.035\pm0.024$ & (1)\\
PNe Type II & $-0.023\pm0.005$ & (1)\\
PNe Type III & $-0.011\pm0.013$ & (1)\\
\Hiiregs\ & $-0.044\pm0.010$ & (2)\\
\enddata
\null (1) \citet{Stanghellini2010} (2) \citet{Esteban2005}.
\end{deluxetable}

\section{Why are the observed \tsqd\ values in gaseous nebulae so large?}\label{sec:t2_large}

To determine abundance ratios, $T_e(4363/5007)$ has been used very often under the assumption of constant temperature ($t^2 = 0.00$). In the presence of temperature inhomogeneities however, the use of $T_e(4363/5007)$ yields oxygen abundances that are smaller than the real ones \citep[e.g.][]{Peimbert1967,Peimbert1969}. In general, under the assumption of $t^2$ = 0.00, the abundance ratios derived from the ratio of a collisionally excited line to a hydrogen recombination line are underestimated, while those derived from the ratio of two collisionally excited lines with similar excitation energies or from the ratio of two RLs, are almost independent of $t^2$.  Nevertheless for some applications, like the determination of the primordial helium abundance, which is based on the ratio of hydrogen and helium RLs, the errors introduced by adopting a $t^2 = 0.00$ value are small but non-negligible \citep[e.g.][]{Peimbert2016}.

From 20 well observed PNe it has been found that $0.024 \le t^2 \le 0.128$, with an average value of 0.064 \citep{Peimbert2014}. While from 37 well observed \Hiiregs\ it has been found that $0.019 \le t^2 \le 0.120$, with an average value of 0.044 \citep{Peimbert2012}. Moreover, \citet{ToribioSanCipriano2017} from 5 \Hiiregs\ in the LMC have  found that $0.028 \le t^2 \le 0.069$ with an average value of 0.038, and from four \Hiiregs\ in the SMC found that $0.075 \le t^2 \leq 0.107$ with an average value of 0.089. 

The majority of the best $t^2$ determinations have been obtained from the ratio of the \ion{O}{2} to [\ion{O}{3}] line intensities in the visual region, see section 3.5.4. It is important to  corroborate these high $t^2$ values by comparing them with other $t^2$ values based on other methods. The following $t^2$ determinations are in agreement with those derived from the \ion{O}{2}/[\ion{O}{3}] method: a) Balmer continuum to H$\beta$ ratio temperatures together with forbidden line temperatures \citep{Peimbert1967, Peimbert1969, Liu1993},  b)  4363/5007 temperatures together with 5007/52 micron temperatures \citep{Dinerstein1985}, c) a high spatial resolution map of the columnar electron temperature in the Orion nebula taken with the Hubble Space Telescope \citep{ODell2003}, d) RLs of  O$^{+}$, O$^{++}$, C$^{++}$, and Ne$^{++}$  together with CELs of these ions \citep{Esteban2004} , e) the \ion{He}{1} vs [\ion{O}{3}] lines method by \citet{Peimbert2012} together with the atomic physics of \citep{Porter2013} \citep{Peimbert2016, Espiritu2017}.

On the other hand typical photoionization models of static \Hiiregs\ chemically homogeneous and with constant density, like those derived with {\sc cloudy} \citep{Ferland2013}, predict $0.000 \le t^2 \le 0.015$ values, with a typical value of about 0.004. 

The difference between the  predicted $t^2$  values based on photoionization models and the observed $t^2$ values  is an open problem and at least ten different possible explanations have been presented in the literature. We will say a few words about them \citep[see also the review by][]{TorresPeimbert2003}.

\subsection{Electron velocity $\kappa$ distributions}

As mentioned in section 4.2 the electron velocity distributions in \Hiiregs\ and PNe are very close to maxwellian ruling out the possibility that $\kappa$ distributions can account for the large ADF values observed \citep{Ferland2016}. Therefore other causes should be found to explain the $t^2$/ADF problem.

Nevertheless $\kappa$ distributions have been used in many papers on gaseous nebulae, both estimating them from observations or computing atomic data for different $\kappa$ values. The $\kappa$ distributions can be represented by temperature inhomogeneities; the $t^2$ needed can be obtained by the following expression:
\begin{equation}
t^2 = \frac{0.96}{\kappa},
\end{equation}
therefore values derived from observations or used in atomic physics computations, in the of $10 \lesssim \kappa \lesssim 50$ range, correspond to the observed values of $0.020 \lesssim t^2 \lesssim 0.100$ \citep{Peimbert2013, Ferland2016}.

\subsection{Chemical inhomogeneities}

\citet{Tsamis2003, Tsamis2005, Stasinska2007} suggested the presence of metal rich droplets produced by supernova ejecta as predicted in the scenario of \citet{TenorioTagle1996}, these droplets do not get fully mixed  with the interstellar medium until they become photoionized in \Hiiregs, these droplets could be responsible for the ADF problem in \Hiiregs. Moreover these droplets, if present, had to be denser and cooler than the surrounding material in \Hiiregs. Based on high quality observations of multiplet V1 of \ion{O}{2} of 8 galactic \Hiiregs\ and one extragalactic \ion{H}{2} region \citet{Peimbert2013} find that the signature of oxygen-rich droplets of high density and low temperature is absent, ruling out the possibility of chemical inhomogeneities in these \Hiiregs.

Similarly \citet{Peimbert2014}, based on high quality observations of multiplet V1 of \ion{O}{2} in 16 galactic planetary nebulae, find ADF values in the 1.42 to 9.67 range (including NGC 6153 with an ADF of 9.67). They also find that there is no sign of temperature low, density high droplets, implying that these objects are chemically homogeneous. For chemically inhomogeneous PNe  the relatively metal rich regions are expected to be cooler than the relatively metal poor ones.

Alternatively \citet{Corradi2015, GarciaRojas2016, GarciaRojas2016a, Jones2016, McNabb2016, GarciaRojas2017, Wesson2017}, and references therein, have found  a group of PNe that show ADF values in the 10 to 80 range. This group includes Haro 2-33 (Hen 2-283), Fg 1, NGC 6778, NGC 6337, NGC 6337, Pe 1-9, MPA 1759, M1-42, Hf 2-2, Abell 63, Abell 46, and Abell 30. This group corresponds to binary stars that after a first ejection of gas produce a second one with a very small mass but with a very high overabundance of O/H relative to the first ejecta. The ADF in these objects is strongly centrally peaked. They propose that these large ADF values should be explained in the framework of close binary evolution, and discuss the possibility that these systems could have gone through a common envelope phase. 

Some giant extragalactic \Hiiregs\ include WR stars and SNe remnants that will produce chemical inhomogeneities, like NGC 2363 \citep{GonzalezDelgado1994}.

\subsection{Deposition of mechanical energy (shocks and turbulence)}

\citet{Peimbert1991} suggest that the high $t^2$ values observed in PNe and \Hiiregs\ might be due to the presence of shock waves. 

\citet{Peimbert1995} find that $T$(C$^{++}$) the temperature derived from the \ion{C}{3} $I$(1906+ 1909)/$I$(4267) ratio is in general considerably smaller than the $T$(O$^{++}$) temperature derived from the $I$(5007)/$I$(4363) ratio, they find that the objects with highest $T({\rm O}^{++})-T({\rm C}^{++})$ values are those that show large velocities and complex velocity fields, and consequently suggest that the deposition of mechanical energy by the stellar winds of the PNe is the main responsible for the temperature differences.

\citet{GonzalezDelgado1994} propose that stellar winds from WR stars play an important role explaining the high $t^2$ values derived for the giant extragalactic \ion{H}{2} region NGC 2363.

\citet{Stasinska1999} find that photoionization models of I Zw 18 yield too low [\ion{O}{3}]$\lambda\lambda$ 4363/5007 ratio by about 30\%, relative to the observed value, the difference might be due to the presence of shock waves and turbulence. Similar results were obtained by \citet{GonzalezDelgado1994} and \citet{Luridiana1999} for NGC 2363, \citet{Luridiana2001} for NGC 5461, and \citet{Relano2002} for NGC 346.

\citet{ODell2015} discuss the presence of stellar outflows in the Orion Nebula. Outflows can produce shocks and turbulence. \citet{Arthur2016} discuss the presence of turbulence in the Orion Nebula. The Orion Nebula is an \ion{H}{2} region ionized mainly by an O7 star, while many, of the giant extragalactic \Hiiregs, for example NGC 5471 \citep{Skillman1985}, I Zw  18 \citep{Skillman1993} and NGC 2363 \citep{GonzalezDelgado1994}, present evidence of large velocity winds probably produced by Wolf-Rayet stars and supernovae. These powerful outflows might be partly responsible for the higher $t^2$ values derived for  these objects than those observed in the Orion Nebula.

\subsection{Time dependent ionization, might be important for PNe but not for \Hiiregs.}

When a photoionization front passes through a nebula it heats the gas above the steady state value and some time is needed to reach thermal equilibrium. When the stellar ionizing flux decreases the outer regions of the nebula become isolated from the stellar radiation field and will continue cooling before fully recombining, creating cold partially ionized outer regions (this might be the case in NGC 7009).

\subsection{Density Inhomogeneities}

Extreme density inhomogeneities  are present in most PNe as can be seen from optical images. Density inhomogeneities produce temperature inhomogeneities, because when the density is high enough, CELs can be de-excited collisionally reducing the cooling efficiency of the forbidden lines.

\citet{Viegas1994} have discussed this possibility and find that for typical, chemically homogeneous, PNe this effect is of the order of $t^2 = 0.005$ or less. For chemically inhomogeneous PNe the effect of density inhomogeneities is crucial. 
 
For steady state photoionization models moderate density inhomogeneities are not very important, however for time dependent processes, the regions of higher density will reach equilibrium sooner than those of lower density.

The relevance of the density inhomogeneities can be estimated from the filling factor (see equation 14), which is almost always considerably smaller than one \citep[e.g.,][]{Mallik1988}.

\subsection{Shadowed regions}

\citet{Mathis1976} proposed the ionization by diffuse radiation of shadowed regions, those regions ionized by photons not directly coming from the ionizing star. The shadowed regions in PNe are expected to have a temperature lower than that of the directly ionized regions from the central star. For the Helix Nebula, NGC~7293,  \citet{Canto1998} and  \citet{ODell2005} estimate  the temperature of the shadowed regions to be from one half to two thirds that of regions ionized directly by the central star. This process is present but it is not the main one producing temperature inhomogeneities. For NGC 7293  the high density knots cover only about 0.05 of the total solid angle (covering factor) as seen from the central star; this could contribute to $t^2$ with about 0.005-0.010.

\subsection{Cosmic rays}

Based on the very small upper limit that they obtain for the $I$(4686, He$^{++}$)/$I$(H$\beta$) ratio, \citet{Peimbert1972} conclude that cosmic rays are not important in the heating rate of the Orion nebula; they also suggest that the observed $I$(4686)/$I$(H$\beta$) ratio in some extragalactic \Hiiregs\ might be due to cosmic ray ionization. \citet{Giammanco2005} propose the ionization by cosmic rays as a key mechanism acting in giant \Hiiregs.
 
\subsection{Spatially distributed ionization sources inside \Hiiregs}

\citet{Ercolano2007} study the effects of spatially distributed ionization sources on the temperature structure of \Hiiregs. They find that  metallicity indicators calibrated by grids of spherically symmetric photoionization models may suffer a systematic bias due to their dependence on the ionization parameter of the system. The O/H values derived by not considering this effect produce underestimates in the 0.1 to 0.3 dex range. These errors are likely to represent the worst case scenario, but nevertheless their magnitude and systematic nature do not allow them to be ignored.

\subsection{Overestimation of the intensity of weak emission lines}

\citet{Rola1994} have shown that the intensity of a weak line with a signal to noise ratio smaller than five is overestimated. This effect might be present in the observations of oxygen and carbon RLs that in a given object are typically three orders of magnitude fainter than the brighter RLs of H. This intensity overestimation could also be present in the [\ion{O}{3}] 4363 line that is typically two or three orders of magnitude fainter than the [\ion{O}{3}] 5007 line. The overestimation of the oxygen RLs produce higher O/H abundances than the real ones, while the overestimation of [\ion{O}{3}] 4363 produces higher \temp\ and lower O/H abundances than the real ones. These two results increase spuriously the ADF values. To avoid this possibility it is important to measure line intensities with a higher than 5 signal to noise ratio.

\subsection{Magnetic reconnection}

\citet{GarciaSegura2001} propose that the multiple, regularly spaced concentric shells around some PNe could be due to the effects of a solar-like magnetic, cycle with a periodic polarity inversion in the slow wind of an asymptotic giant branch (AGB) star. Presumably these shells of alternating polarity could give rise to magnetic reconnection processes once that they are compressed in the formed PN (i. e. in the swept-up shell). Magnetic fields have been measured in the torus surrounding K3-35 \citep{Miranda2001}; also the presence of a magnetic field has been inferred in the PN OH0.9 + 1.3 \citep{Zijlstra1989} and in the pre-planetary nebula IRAS 17150-3224 \citep{Hu1993}.

\subsection{Summary on $t^2$}

From the previous discussion we consider that the best abundance determinations are those derived from the oxygen and hydrogen RLs and that the $t^2$ values derived from observations should be used to estimate the abundances derived from forbidden lines of the other elements. 

\Hiiregs\ have $t^2$ values in the 0.02 to 0.12 range, with a typical value of 0.045, and ADF values in the 1.5 to 3 range with a typical value of 2. We conclude that probably in \Hiiregs\ the main source of temperature inhomogeneities is due to the deposition of mechanical energy by shocks and turbulence produced by stellar winds, including those winds due to SNe and WR stars in giant extragalactic \ion{H}{2}  regions. 

We divide PNe in two groups: a) classical PNe with homogeneous chemical composition and b) PNe with inhomogeneous chemical composition.

Classical PNe have the following properties: a) they do not have gaseous inclusions of high density and low temperature, b) they have $t^2$ values in the 0.024 to 0.128 range with a typical $t^2$ value of 0.065, c) they have ADF values in te 1.4 to 10.0 range with a typical value of 2.3, and d) they have typical ionized masses in the 0.05 to 0.40 M$_{\odot}$ range \citep[e.g.][]{Mallik1988}. 

PNe with inhomogeneous chemical composition have the following properties: a) they do have inclusions of high density and low temperature in the gaseous envelope, possibly the result of close binary evolution, b) their O/H values decrease from the central star outwards, c) they present large temperature inhomogeneities that cannot be analyzed with the $t^2$ formalism in a one phase model, d) their ADF values are in the 10 to 100 range and e) they have typical ionized masses of about 0.01M$_{\odot}$.

For chemically homogeneous PNe the best abundance determinations are those derived from oxygen and hydrogen RLs. We also conclude that probably the main source of temperature inhomogeneities in PNe is due to the deposition of mechanical energy from strong stellar winds. 

For chemical inhomogeneous PNe, the best approach to determine the chemical abundances is to construct two phase models, but still a particular kind of $t^2$ should be considered when fitting such models: by their very essence such models will not have a homogeneous temperature, moreover, with inhomogeneous chemistry the $t^2$ value can be very different for each chemical element and for each ion, most of this will be directly considered when studying the two phases of the model; however, the hot phase alone is expected to behave similarly to a chemically homogeneous PN, and as such it is expected to have its own particular $t^2$ (and ADFs); since most of the CEL intensities are expected to come from this phase, the $t^2$ correspondending to the hot phase alone is necessary and should be sufficient to reproduce the behavior of all the optical CELs observed in these objects.\\

\section{Final remarks}
The determination of accurate chemical abundances of PNe and \Hiiregs\ is a fundamental problem in astrophysics. The chemical abundances of PNe provide us with constrains for stellar models in the 0.8 M$_\odot$ to 8 M$_\odot$ range. \Hiiregs\ permit us to determine the initial abundances with which stars are formed. Both sets of data are fundamental for the study of stellar and galactic chemical evolution.

Also accurate abundances of \Hiiregs\ and PNe are needed to estimate the fraction of the elements embedded in dust grains. These fractions are needed to study the processes of formation and destruction of dust in gaseous nebulae.

Further efforts are needed to obtain data for \Hiiregs\  and PNe in galaxies of different types. In spiral galaxies observations of \Hiiregs\  at different galactocentric distances to study which has been the evolution of the ISM as a function of time for these galaxies. The study of very metal poor extragalactic \Hiiregs\  will permit to increase the accuracy of the primordial helium determination, this value is important for cosmology and particle physics.

The ADF problem should be studied further to obtain a consensus on its origin and scope. The same can be said for the high $t^2$ values derived from observations of \Hiiregs\ and PNe.

\acknowledgments
We thank the referee for many excellent suggestions. We are grateful to all those that have worked and are working in this field, they have inspired us over the years. MP, AP, and GD-I received partial support from CONACyT grant 241732. MP and AP acknowledge support from PAPIIT (DGAPA-UNAM) grant no. IN-109716. GD-I acknowledges support from PAPIIT (DGAPA-UNAM) grant no. IA-101517.

\allauthors
\listofchanges

\end{document}